\documentclass[useAMS,usenatbib]{mn2e}
\usepackage{graphicx}
\usepackage{amssymb}
\usepackage{natbib}
\usepackage{muni}

\newcommand{\NCR}{{\tt{NCR}}}
\newcommand{\DCR}{{\tt{DCR}}}
\newcommand{\LCR}{{\tt{LCR}}}
\newcommand{\MCR}{{\tt{MCR}}} 
\newcommand{\HCR}{{\tt{HCR}}}
\newcommand{\HHCR}{{\tt{Hi-HCR}}}

\begin{document}

\title{Role of Cosmic Rays in the Circumgalactic Medium}


\author[Salem, Bryan \& Corlies]{Munier Salem$^{1}$, Greg L. Bryan$^{1}$, and Lauren Corlies$^{1}$ \\
$^{1}$Department of Astronomy, Columbia University, 550 West 120th Street, New York, NY 10027, USA}

\date{}

\maketitle


\begin{abstract}
We explore the impact of cosmic rays (CRs) on cosmological adaptive-mesh refinement simulations of a forming $10^{12} M_\odot$ halo, focusing on the circumgalactic medium (CGM), and its resulting low-redshift structure and composition. In contrast to a run with star formation and energetic feedback but no CRs, the CR-inclusive runs feature a CGM substantially enriched with CRs and with metals to roughly $0.1Z_\odot$, thanks to robust, persistent outflows from the disk. The CR-inclusive CGMs also feature more diffuse gas at lower temperatures, down to $10^4$ K, than the non-CR run, with diffuse material often receiving a majority of its pressure support from the CR proton fluid. We compare to recent observations of the CGM of $L\sim L^\star$ galaxies at low redshift, including UV absorption lines within background quasar spectra. The combination of metal-enriched, CR-driven winds and large swaths of CR pressure-supported, cooler diffuse gas leads to a CGM that provides a better match to data from COS-Halos (for HI, SiIV, CIII and OVI) than the non-CR run. We also compare our models to recent, preliminary observations of diffuse gamma-ray emission in local group halos. For our lower CR-diffusion runs with $\kappa_{\rm CR} \in \{0.3,1\} \times 10^{28}$ cm$^2$/s, the CR enriched CGM produces an inconsistently high level of gamma emission. But the model with a relatively high $\kappa_{\rm CR} = 3 \times 10^{28} {\rm cm^2/s}$ provided a gamma-ray luminosity consistent with the ``extra-galactic'' gamma-ray background observed by FERMI and roughly consistent with preliminary measures of the emission from M31's CGM.
\end{abstract}

\begin{keywords}
galaxies:formation - CGM: cosmic rays - methods:numerical
\end{keywords}

\section{Introduction}

The Circumgalactic Medium (CGM) denotes the diffuse, multiphase medium that fills galaxy halos beyond their central, star-forming (SF) regions. Inflows from the pristine intergalactic medium \citep[IGM; e.g.][]{Binney1977,Birnboim2003,Keres2005,Keres2009, Ocvirk2008, vandeVoort2011} and enriched outflows from the SF disk flow through this interface, endowing it with mass and (in the latter case) metals and altering its temperature and kinematic structure \citep[e.g.][]{Cowie1995,Porciani2005,Mandelbaum2006,Oppenheimer2008,Dave2012,Behroozi2010, Hummels2013,Ford2014, Ford2015, Suresh2015}. Studying the CGM thus provides a sensitive probe into the process of galaxy formation.

Observations in absorption have begun to characterize the tenuous CGM, both at high redshift \citep[e.g.][]{Simcoe2006, Steidel2010} and in the local universe in X-ray \citep{Anderson2010,Gupta2012,Miller2013, Miller2015} and UV \citep[e.g.][]{Morris1993}. Recent work suggests the CGM of an $\sim L^*$ galaxy may host a majority of the halo's baryons and metals, an appreciable fraction of which may be in a relatively cool, $<10^5$ K, state \citep[e.g.][]{Tumlinson2011,Werk2013,Werk2014}. 

Galaxy simulations have employed a variety of models to capture the energetic feedback that regulates star formation and produces observed thin disks with flat rotation curves \citep[e.g.][]{Governato2010,Brooks2011,Guedes2011,Agertz2013,Aumer2013,Stinson2013,Marinacci2014}. A recent approach has been the inclusion of cosmic-ray protons (CRs) via a two-fluid model \citep{Ensslin2007,Jubelgas2008,Wadepuhl2011,Uhlig2012} which has recently proven capable of launching winds and regulating star formation in an $M \sim 10^{12} M_\odot$ halo \citep{Salem2014a,Salem2014b}.

In this paper, we analyze a suite of simulations of a forming $10^{12} M_\odot$ galaxy that include a diffusive CR proton fluid, focusing on how CRs impact the halo's low-redshift CGM. This work is a direct follow-on to Salem et al 2014 (hereafter SBH14), in which we focused on the impact of CR protons on galactic disk structure and star formation. In section \ref{sec:cgm-methodology} we provide a brief description of the simulations and CR model. In section \ref{sec:cgm-structure-comp} we explore the structure and composition of the CGM across runs with disparate CR diffusion coefficients, highlighting the total mass and temperature structure of our CGM.  In section \ref{sec:cgm-metal-obs} we produce column maps of ion species in the CGM and compare to the recent Cosmic Origins Spectrograph Halos Survey (COS-Halos) results. In section \ref{sec:cgm-metal-obs} we compute the gamma-ray luminosity of hadronic losses in our CR-infused halo, comparing to recent analysis of Fermi LAT data. In Section \ref{sec:cgm-discussion} we explore how our analysis constrains models of CR diffusion within the CGM and discuss implications on the total baryon budget of the CGM of MW-sized galaxies. Finally, Section \ref{sec:cgm-conclusion} summarizes the results of this paper.

\section{Methodology}
\label{sec:cgm-methodology}

In this section we provide a brief summary of the simulations analyzed in this work. This paper presents a CGM-focused analysis of the same simulations described in \cite{Salem2014b}, hereafter SBH14, which used the adaptive mesh refinement (AMR) hydrodynamics code \verb|Enzo| \citep{Bryan2013}. For a detailed description of the initial conditions and gas physics involved we refer the reader to \cite{Hummels2012} and \cite{Hummels2013} (hereafter HB12 and H13, respectively), which provided analysis of an identical forming halo, though sans-CRs. For a detailed  description of the cosmic ray physics and numerical methodology therein, we refer the reader to \cite{Salem2014a}, hereafter SB14.


\stdTable{Parameters}{
\begin{tabular}{ll}
\toprule
 \multicolumn{2}{c}{\textbf{CR Physics}} 							\\
$\kappa_\crs$		&	$\{0,.3,1,3\} \times10^{28}$ cm$^2$/s		\\
$\gamma_\crs$		& 	$4/3$								\\
$c_{s, {\rm max}}$	&	$1000$ km/s							\\
\midrule
\multicolumn{2}{c}{\textbf{SF / Feedback}}			\\
$\epsilon_{\rm SF}$		& .01					\\
$\delta_{\rm SF}$		& $10^3$				\\
$\epsilon_{\rm SN}$ 		& $3\times10^{-6}$		\\
$y$					& .02					\\
$f_\crs$				& 0.3					\\
\midrule
\multicolumn{2}{c}{\textbf{Cosmology}}		\\
$\Omega_{0}$				& 0.258		\\
$\Omega_{\rm \Lambda}$		& 0.742		\\
$\Omega_{\rm b}$			& 0.044		\\
$h$ 						& 0.719		\\
$\sigma_8$				& 0.796		\\
$z_i$					& 99 			\\
\midrule
\multicolumn{2}{c}{\textbf{Numerics}}				\\
$\Delta x_{\rm min}$			&	425 pc*	\\
size						&	20 Mpc*	\\
\bottomrule
\label{tab:params}
\end{tabular}
}{Simulation parameters}

\subsection{Simulation Physics}
\label{sec:cgm-sim-physics}

\subsubsection{Initial Conditions / Cosmology}
We simulate a $(20 \; {\rm Mpc}/h)^3$ box at a base resolution of $128^3$ cells with WMAP 5-year cosmological parameters \citep{Komatsu2009} from $z=99 \to 0$. We then selected a $1.2 \times 10^{12} M_\odot$, relatively isolated halo of interest (dubbed halo 26 in H12). Within this box we identified the rectangular region spanned by all DM particles that contribute to this halo of interest, and provided two additional levels of static mesh refinement, each a factor of 2 finer resolution in each dimension (providing a factor of 64 more DM particles, and thus a DM particle mass of $4.9 \times 10^6 M_\odot$) to perform a ``zoom simulation'' of the halo. For the baryonic components within this region, we provide an additional 7 levels of AMR to strategically allocate resolution where the gas dynamics grow complex, providing a maximum comoving resolution of 305 $h^{-1}$ pc within dense, star-forming regions. Table \ref{tab:params} summarizes important parameters of these simulations.

\subsubsection{Cosmic Ray Physics}
To capture the dynamical effects of cosmic ray protons, we employ a two-fluid model for the gas and CRs \citep{Drury1986,Drury1985,Jun1994}. The model assumes a relativistic population of $\sim$GeV protons which obey the ideal gas law with $\gamma_\crs = 4/3$. SB14 found the exact choice of $\gamma_\crs$ had little effect on the outcome of the simulations. The CRs streaming velocity is considerably less than the speed of light, $c$, since the ISM's inhomogenous magnetic field scatters their motion.  This effectively freezes the CR fluid to the thermal plasma, where it advects with the bulk motion of the thermal ISM, undergoing adiabatic expansion and contraction, exerting a scalar pressure on the gas, and thus exchanging energy with the thermal ISM via the momentum equation. For the redistribution of CR energy density among fluid parcels that does occur, we adopt a simple scalar diffusion model with a space- and time-homogenous coefficient $\kappa_{\rm cr}$. We neglect non-adiabtic CR energy loss terms, anisotropic diffusion along magnetic field lines, and diffusion of the CRs in energy. This model also does not track the CR's distribution in momentum space (this is explored in Appendix A). \verb|Enzo| solves the two-fluid equations with the fast, robust ZEUS-hydro~\citep{Stone1992}, a space-centered, time-forward, flux conservative, second-order finite difference scheme.

SB14 found the choice of $\kappa_{\rm CR}$ could greatly enhance or diminish the presence of mass-loaded winds above star forming regions of an idealized disk galaxy. In particular, lower diffusion coefficients produced heavier mass-loading, while higher coefficients had more rarefied winds. Ultimately these winds substantially reduced star formation in the disk, especially for lower diffusion coefficients. SBH14 found similar results for SF in a cosmological setting, while also noting the power of the specific CR model in shaping the low-redshift disk structure: higher diffusion runs, particularly $\kappa_\crs = 3 \times 10^{28}$ cm$^2$/s, produced thinner, more extended stellar disks, with more cold gas remaining at low redshift, enhanced spiral arm structure, and flatter (though still far too cuspy) rotation curves. A primary aim of the present work is to once again vary the CR diffusion coefficient, with $\kappa_\crs \in [ 0 , 3\times 10^{27}, 10^{28}, 3 \times 10^{28} ]$ cm$^2$/s to judge its effect on properties of the CGM. CR propagation models and observations of the high energy ISM suggest $\kappa_\crs \sim 10^{28}$ cm$^2$/s \citep[e.g.,][]{Strong1998,Ptuskin2006, Ackermann2012,Tabatabaei2013}. A more accurate model would treat this diffusion process via an anisotropic tensor, whose components would depend on an explicitly modeled magnetic field and the momentum distribution of CRs. However, such an approach requires a spatial resolution inaccessible to the present generation of cosmological simulations.

\subsubsection{Star Formation and Feedback}
Our maximum comoving cell-size of $305 \;h^{-1} \; {\rm pc}$ does not resolve the dense molecular clouds within which individual stars form and die in violent supernovae. To capture the physics of star formation, we thus resort to collisionless ``star particles'' to represent clusters of stars with mass $M_\star \ge 10^5 M_\odot$. The formation rate of these particles is determined via the procedure of \cite{Cen1992}, with updates from \cite{OShea2004}, who adopted a star formation rate (SFR) of $\dot{\rho_{\rm SF}} = \epsilon_{\rm SF} \rho / t_{\rm dyn}$, where $t_{\rm dyn}$ denotes the dynamical time and $\epsilon_{\rm SF}$ the SF efficiency. Briefly, a star particle is created when the cell's gas density relative to the maximally refined sub-grid's base density, $\delta_{\rm SF}$, exceeds a chosen threshold; the velocity field surrounding the cell is convergent (i.e. the local divergence is negative);  and $t_{\rm cool} < t_{\rm dyn}$ within the cell. H12 calibrated $\epsilon_{\rm SF} = .01$ and $\delta_{\rm SF} = 10^3$ to match the Kennicut-Schmidt relation.

The simulations presented here also include feedback from Type II supernovae, by returning energy, mass and metals to the ISM over a dynamical time within SF regions. HB12 explored the impact on star formation of various choices for the energy efficiency, $\epsilon_{\rm SN} \sim 10^{-5}$, the fraction of the star particle's rest energy returned to the gas field. Realistic choices for this quantity (with CRs) failed to strongly suppress SF or redistribute baryons within the galactic disk, though suppression of gas cooling within SF regions had some success lowering the peak rotation curve. H13 also explored the impact of various feedback models on the CGM, and found that again cooling suppression was most successful in delivering sufficient metals to the CGM to match low-redshift quasar absorption lines (although no model had much success matching the warm-hot gas absorber OVI). Here we adopt the most modest $\epsilon_{\rm SF} = 3 \times 10^{-6}$ explored in those works (corresponding to $10^{51}$ ergs for every 185 $M_\odot$ of stars formed).  As in that work, the mass reintroduced to the ISM is metal-enriched with a yield $y = .02$. For all but our non-CR physics simulation, we also divert a fraction $f_{\rm CR} = 0.3$ of this energy feedback into the relativistic CR fluid (the entirety of mass and metal feedback is deposited into the thermal fluid, as the CR's mass is negligible). SB14 explored the importance of $f_{\rm CR} \in [0.0,0.3,1.0]$ and found that while larger values enhanced the mass-loaded winds driven from SF regions, all CR-feedback inclusive runs had the same basic picture of robust winds and suppressed SF. We do \emph{not} implement cooling suppression or any other \emph{ad hoc} prescription to produce galactic winds, instead relying on the CR physics to naturally drive outflows.

\subsubsection{Chemistry, Heating and Cooling}
We explicitly track the balance of all ion species of hydrogen and helium via a non-equilibrium primordial chemistry network that includes recombination, bremsstrahlung radiation, Compton cooling, collisional ionization, photoionization and photoexcitation \citep{Smith2008,Abel1997,Anninos1997}. For heavier elements, {\tt Enzo} keeps track of a single metallicity field. A lookup table generated via \verb|Cloudy| \citep[version 07.02.01,][]{Ferland1998} synthesizes these chemical abundances and the thermodynamic state of the gas into bulk heating and cooling rates. In addition to radiative cooling, we include an isotropic, homogenous, redshift-dependent UV background that heats the gas \citep{Haardt2001,Haardt1996}. These tracked quantities and the UV background are also used in Section \ref{sec:cgm-metal-obs} to construct ion column density maps in our simulated CGM. Heating and cooling processes were tabulated down to a temperature floor of $10^4$ K.
 
\subsection{Description of Simulation Suite}
\label{sec:cgm-simulation-suite}

\stdTable{Simulation Suite}{
\begin{tabular}{lll}
\toprule
Name	&	Description			& $\kappa_{\rm CR}$ $(10^{28}$ cm$^2$s$^{-1})$ \\
\midrule
\tt{NCR}		&	No Cosmic Rays		&	---		\\
\tt{DCR}		&	Diffusionless 			&	0		\\
\tt{LCR}		&	Less-diffusive 			&	0.3		\\
\tt{MCR}		&	Moderalely-diffusive 		&	1.0		\\
\tt{HCR}		&	Highly-diffusive 		&	3.0		\\
\bottomrule
\label{tab:sim-suite}
\end{tabular}
}{}

Five simulations of the same dark matter (DM) halo are presented here with nearly identical physics except for the cosmic ray model. Table \ref{tab:sim-suite} summarizes each run's distinct features. \verb|NCR| is a run devoid of CRs, where all SN feedback is injected into the thermal ISM (i.e. $f_{\rm CR} = 0.0$). \verb|DCR| is a diffusionless CR run, where the two-fluid model was employed and CRs are injected during SF, but the high-energy CR component is unable to leave the original Lagrangian fluid parcel within which it was created. This run serves mostly to understand the phenomenology of the CR model and creates galaxies with very unusual, unobserved baryonic properties. Thus we exclude this run from many parts of the analysis for clarity. Finally three runs with the full, diffusive CR model are presented, which we refer to as \verb|LCR|, \verb|MCR| and \verb|HCR|, for less-, moderately- and highly-diffusive. These employ $\kappa_{\rm CR} = [ .3, 1, 3] \times 10^{28}\;{\rm cm}^{-3}$, respectively. 

\section{CGM Structure and Composition}
\label{sec:cgm-structure-comp}

In this section we explore the structure and composition of the CGM of our $\sim10^{12} M_\odot$ halo across simulations.  SBH14 explored the central SF region of the same suite of runs, most notably finding that the more diffusive CR-inclusive models had the greatest success producing extended, rotationally-supported disks of cold gas and stars, while less-diffusive CR-inclusive runs produced the best match to observed rotation curves for MW-mass systems. Here we largely ignore the disk and stellar component, instead focusing on the region beyond $10$ kpc of the galactic center out to beyond the virial radius. We also exclude the non-diffusive CR run from the majority of figures and discussion, since its role was mostly to aid in understanding the phenomenology of the CR model. The analysis presented through the remainder of this paper made extensive use of the \verb|yt| analysis software \citep{Turk2011}. 

\subsection{Surface Densities}

\stdFullFig{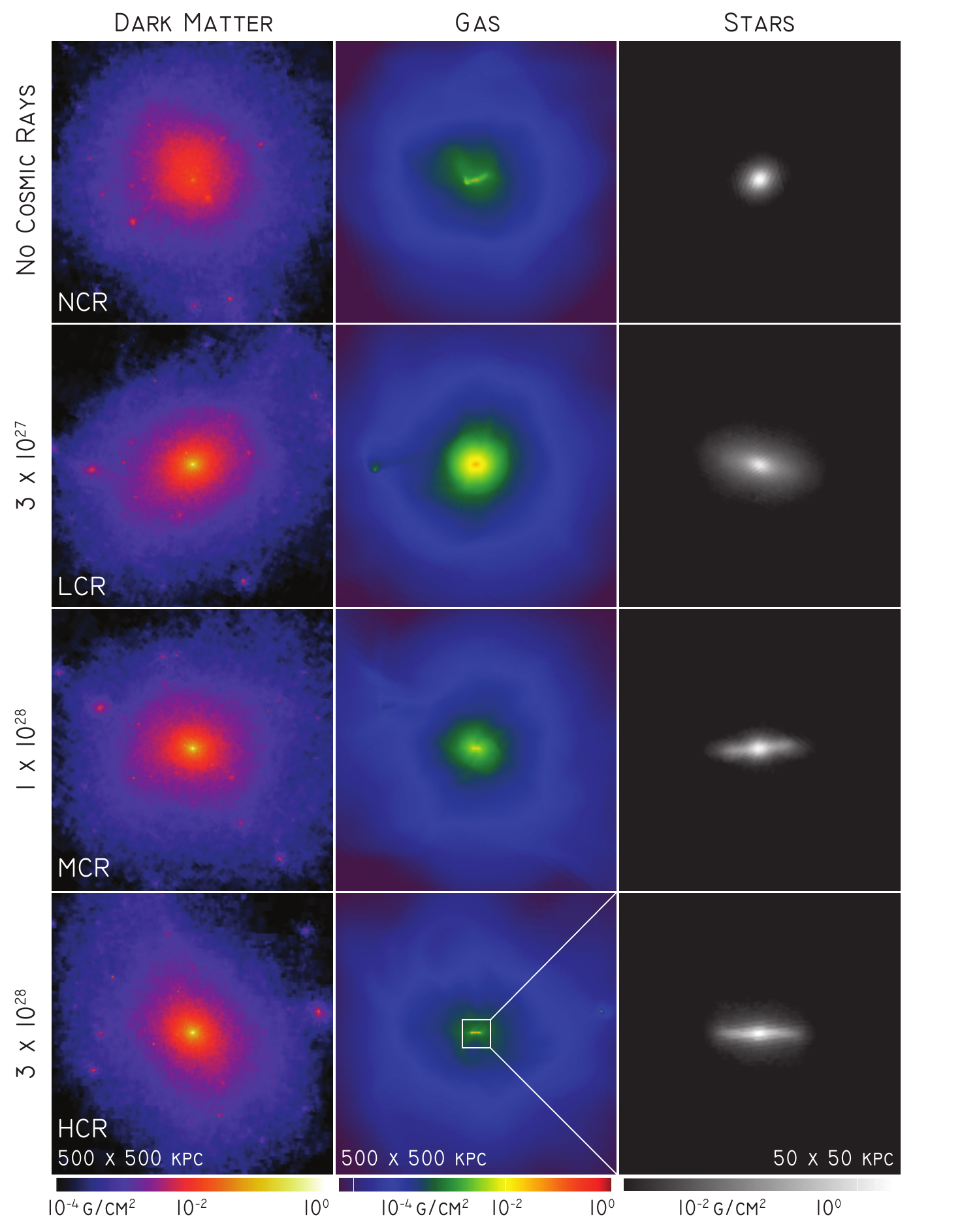}{Density projections (surface density) across our simulations for the three massive components of the runs: dark matter (DM), gas and stars at $z=0$. The extended DM profile appears identical across runs, though runaway SF and feedback in the massive stellar bulge of the no-CR run has heated the DM, reducing its central density peak SBH14. In contrast the diffuse halo gas surrounding the disk and extending tens of kpc is very much dependent on the CR-diffusion model chosen, with lower-diffusion runs exhibiting higher surface densities. The stellar disk of each run, analyzed in SBH14, is shown for reference, with the degree of rotational support increasing as the runs become more diffusive. All projections in this paper are shown from an ``edge-on'' perspective, i.e. the $\hat{y}$-axis of each panel is aligned with the rotationally supported disk's angular momentum axis.}{fig:dm-gas-stars}{.8}

We begin by observing the distribution of mass within the halo. Figure \ref{fig:dm-gas-stars} shows density projections (i.e. surface density) of dark matter (DM), gas and stars across our runs. The extended DM halo appears roughly identical across runs, though runaway star formation (SF) and feedback in the massive stellar bulge of the \NCR\ run has heated the DM, reducing its central density peak. Note that all projections in this work are taken with their $\hat{y}$-axis aligned with the angular momentum vector of the rotationally supported disk. Thus each panel shows the DM profile from a different perspective and satellites appear in different relative locations across runs. From the gas density plots, we find the CR-inclusive runs feature a noticeably higher gas surface density within $\sim 50$ kpc of the galactic center, most notably with the surface density $\sim 10$ kpc above the disk in the \LCR\ run over a decade higher than in \NCR. This enhanced gas surface density is substantially reduced as the CR runs become more diffusive. For reference, a zoomed-in projection of the stellar disks, discussed in SBH14, are shown in the rightmost column of Figure \ref{fig:dm-gas-stars}.

\stdFullFig{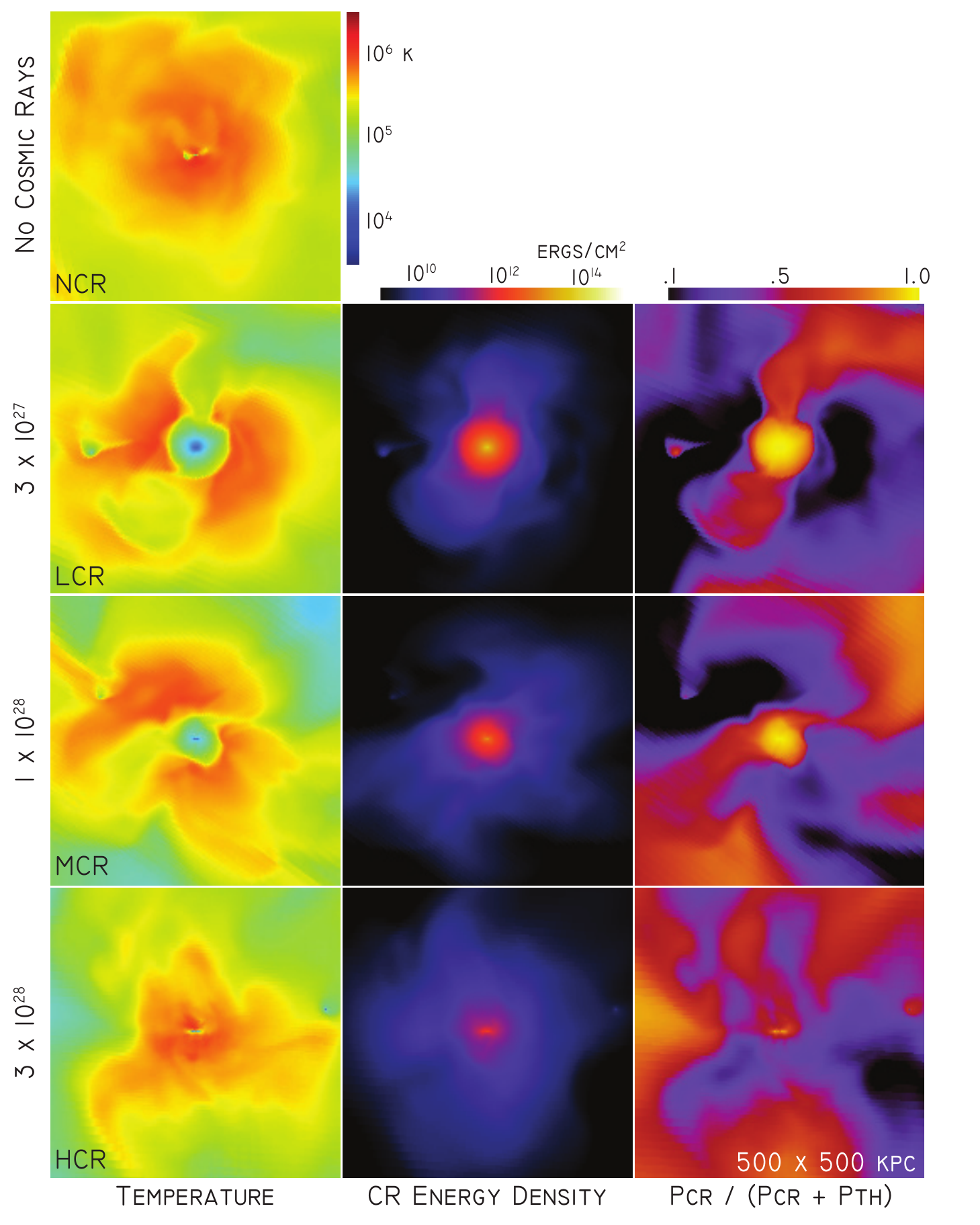}{Gas-density weighted projections of gas temperature, CR energy density, and the ratio of $P_{\rm CR}/(P_{\rm CR} + P_{\rm th})$, a measure of the CR fluid's dynamical dominance, all at $z=0$. While the non-CR run's CGM is dominated by gas at $T \sim 10^6$ K, the CR-inclusive runs all feature a stronger presence of gas below $10^6$ K. For the less diffusive runs, a large pocket of gas $\sim 10^5$ K develops within $\sim 50$ kpc of the galactic center, where CRs provide total pressure support for the diffuse CGM material. Farther away from the SF region, across all CR-inclusive runs, there exists broad swaths of diffuse CGM material between $10^{4.5} - 10^{6}$ K, with hotter portions mostly thermally supported, whereas cooler portions are CR-supported. In short, the CR component allows broad swaths of CGM gas to drop below the temperature required of a purely thermal CGM while still maintaining the hydrostatic balance of the composite fluid.}{fig:temp-crs}{.9}

The inclusion of cosmic ray protons has a profound effect on the temperature structure of our CGM, as demonstrated in Figure \ref{fig:temp-crs}, which shows mass-weighted projections of temperature, $\epsilon_{\rm CR}$, and $P_{\rm CR}/(P_{\rm CR} + P_{\rm th})$, a measure of the CR fluid's dynamical dominance. For \NCR, the CGM out to $\sim50$ kpc is dominated by $T \sim 10^6 $ K gas, with the temperature dropping towards $10^5$ K by the virial radius. When we switch on the CR physics, however, broad swaths of our CGM drop in temperature. For \DCR\ and \MCR, this includes a large pocket of gas below $10^5$ K out to $\sim 50$ kpc, where much of the gas is below $10^{-3} \;{\rm cm}^{-3}$ in density. This relatively diffuse, cold gas can exist in mechanical equilibrium thanks to the strong presence of CR protons in an extended halo about the SF region. This diffuse HI bubble grows less prominent for the more diffusive runs, most notably for \HCR, where the thermal component once again becomes the dominant pressure source within 50 kpc. 

For all CR-diffusion runs, the CGM beyond $\sim 50$ kpc features a broad range of gas temperatures, with thermal pressure-dominated swaths at $\sim10^6$ K as before, as well as large regions below $10^{5}$ K, where the CR fluid provides the bulk of support: the composite fluid now has a second source of pressure support, allowing diffuse gas to drop below the temperature required of a purely thermal CGM while still maintaing hydrostatic balance. This has major implications regarding how well our simulations can reproduce recent observations of metal columns in the CGM of $L \sim L^*$ galaxies at low redshift, which we explore in Section \ref{sec:cgm-metal-obs}. 

For $d \gtrsim 50\;{\rm kpc}$ the diffusion time scale for $\kappa_{\rm CR} \sim 10^{28} \; {\rm cm^2/s}$ is $\gtrsim 1$ Gyr, and thus the CR fluid is effectively non-diffusive at this length-scale.

\stdFig{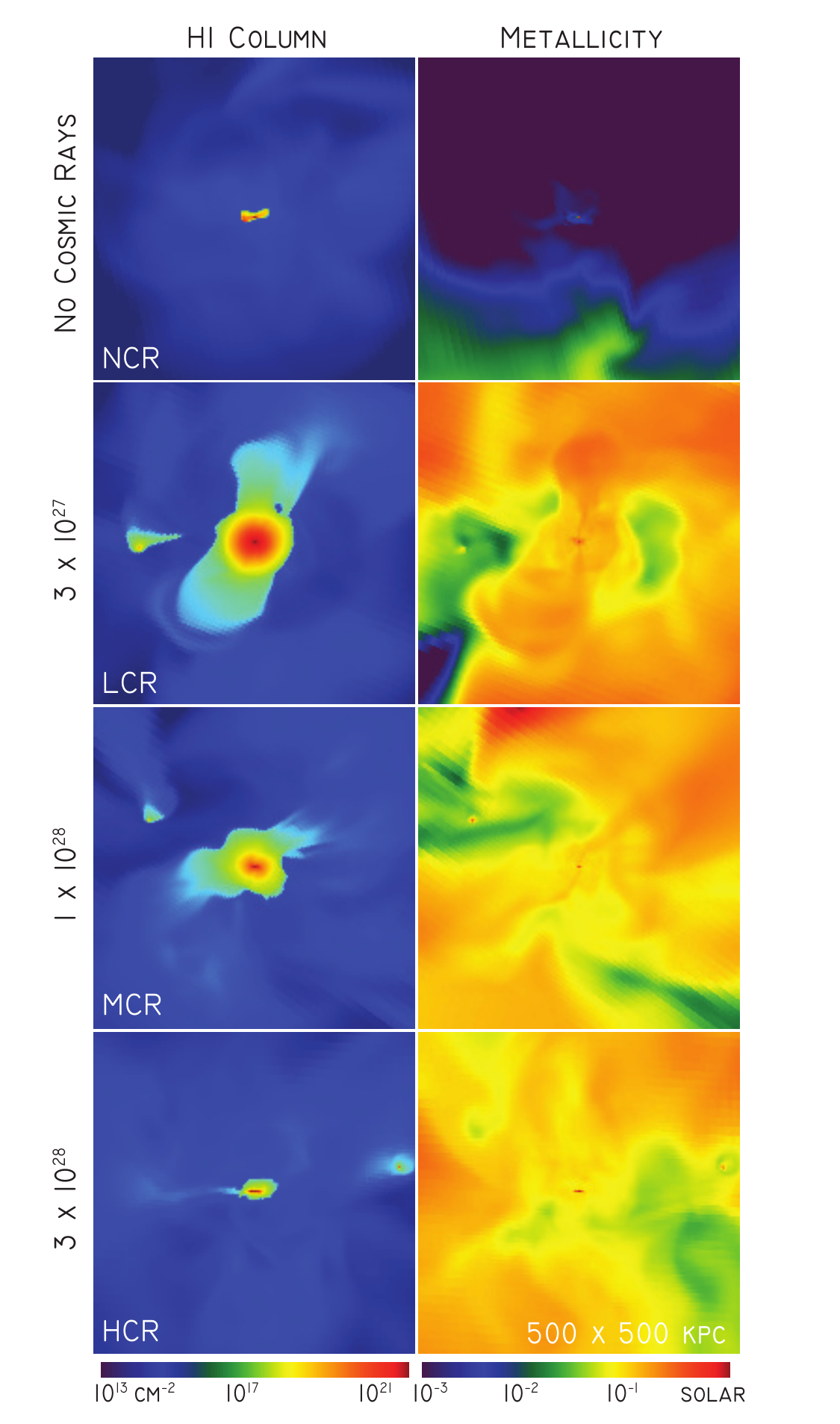}{Column densities of HI and mass-weighted projections of metallicity across simulations at $z=0$. The CR-inclusive runs feature far more HI gas beyond the SF disk, and likewise a far higher metallicity CGM. The covering fraction of HI is strongly dependent on CR model, with the least diffusive runs exhibiting the broadest swath of high column atomic hydrogen, in regions coincident with lower temperatures and higher densities (see Figure \ref{fig:temp-crs}) where the CR-fluid is dominant.}{fig:basic-chem}{.8}

As described in Section \ref{sec:cgm-methodology}, our simulations explicitly track multiple ionization states of both hydrogen and helium, as well as the metal fraction of our thermal gas, allowing a straightforward method of producing the HI column maps and mass-weighted projections of metallicity shown in Figure \ref{fig:basic-chem}.  From the HI column maps, we find neutral hydrogen has a far stronger presence in our CR-inclusive CGMs in regions of higher gas density, lower temperature and higher $\epsilon_{\rm CR}$, which are also CR-pressure supported. This is most pronounced in the lowest diffusion \LCR\ run, but the effect persists to the highest diffusion run. A satellite features a robust HI component in all three CR-inclusive runs shown.

The right column of Figure \ref{fig:basic-chem} shows a mass-weighted projection of metallicity. Although a cloud of enriched material exists beyond the SF region of our non-CR run, the CGM within the virial radius of this halo is largely metal-poor. In contrast, the CGM of our CR runs is metal enriched, with the projections showing $Z \gtrsim 0.1 Z_\odot$ for a majority of pixels. This holds true across diffusion parameters. In particular, \MCR\ and \HCR\ are devoid of a single sightline within the virial radius where the metallicity falls below $10^{-1.5}$. We discuss the dichotomy between the enrichment of non-CR and CR runs in Section \ref{sec:cgm-discussion}.  The lack of metals in the halo in the NCR run is due to the inability of simple thermal energy feedback models in cosmological galaxy simulations to efficiently drive winds because their energy is rapidly radiated away \citep[e.g.,][]{Katz1996, Hummels2012}.  Simulations including CRs produce substantive outflows \citep{Salem2014a}, which can eject metals into the CGM; we also note that other feedback models can drive outflows but discuss the unique signatures of a CR-dominated CGM in Section 6.

\subsection{Radial Profiles}
\label{sec:cgm-radial-profiles}
\stdFullFig{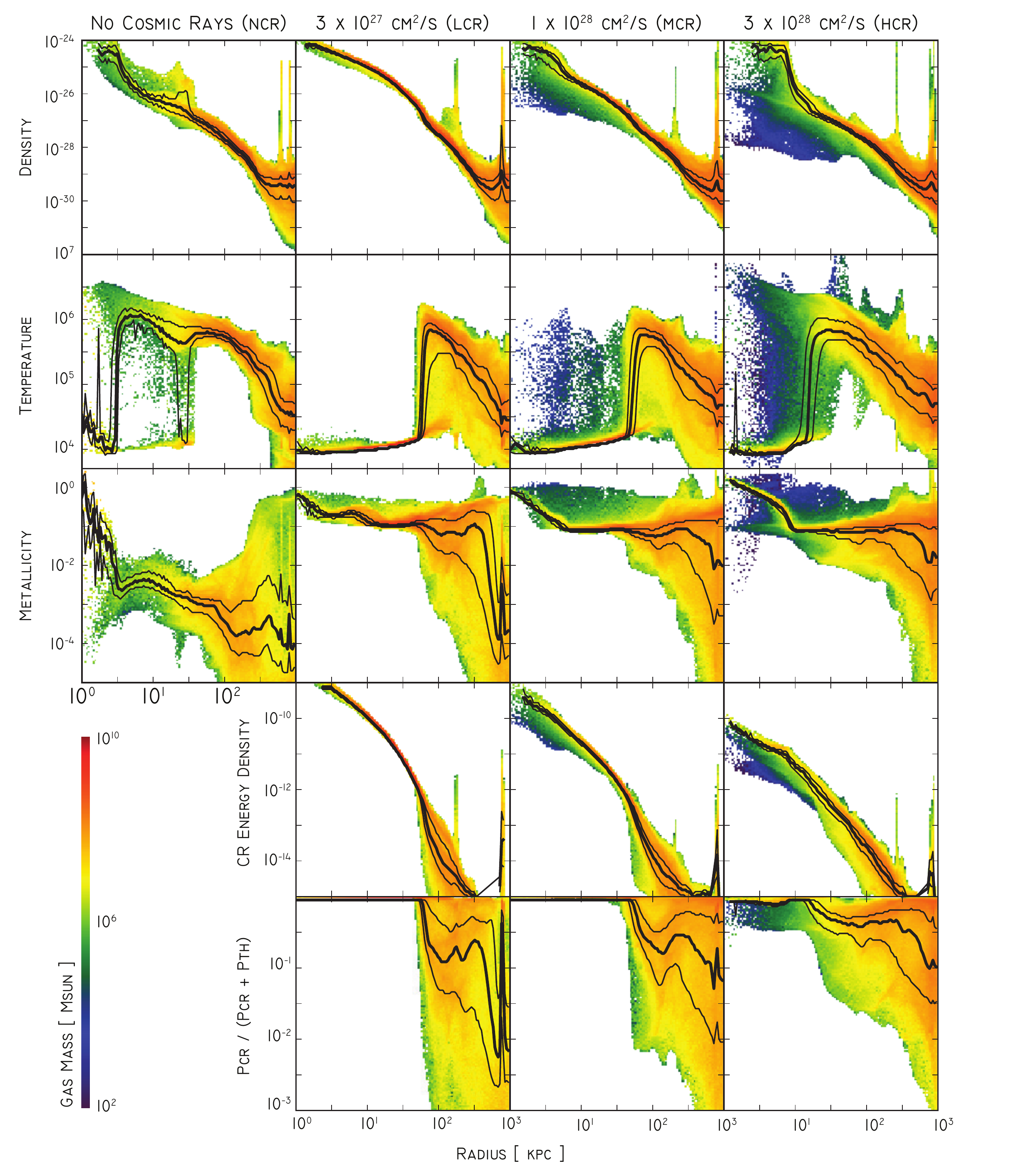}{Profiles of CGM gas properties versus radius: effectively 2D histograms binned by gas mass. Overplotted are median values as a function of radius (thick lines) along with quartile bounds (thin lines). All quantities are plotted at $z=0$. }{fig:profiles}{.6}

Figure \ref{fig:profiles} seeks to quantify the gas properties we've looked at thus far as a function of radius from the galactic center of our halo.  Displayed are two-dimensional histograms, binned by gas mass, of various gas properties versus radius. Also over plotted are lines representing the median and quartile values as a function of radius. From the density profiles (top row) we find the low-diffusion CR runs feature systematically higher gas densities than the non-CR run at each radius, with an extremely tight spread about the median value. For the more-diffusive runs, the spread in density expands at small radii($r < 30$ kpc) where a minority of the gas exists at densities a factor of 100 below the non-CR run's distribution. Across the CR runs, within $\sim 30$ kpc, the gas is almost exclusively cold ($10^4$ K), in stark contrast to the non-CR run's median value $\sim 10^6$ K at these radii. The more diffusive runs do feature a minority of gas up to these temperatures, but the vast majority of gas remains cold in the inner CGM. At large radii the least diffusive runs now show the most spread in temperature. 

The metallicity profiles (third row) corroborate the stark contrast between CR and non-CR runs seen in the earlier projections: within the virial radius, the non-CR CGM is devoid of gas beyond 0.01 solar metallicity. Meanwhile, the CR runs feature a CGM almost exclusively above 0.1 solar. The most diffusive run has the largest spread in metallicity, especially approaching and beyond the virial radius. 

Finally we analyze the CR pressure dominance in the CGM (bottom row). Across CR runs, this quantity is strictly unity (completely CR dominated) within $\sim 30$ kpc for the two less-diffusive runs, and nearly as monolithically CR-dominant for \HCR. However, beyond this radius, all three runs exhibit a broad distribution of the pressure ratio, with the central 50\% of gas parcels anywhere from 50\% CR-pressure supported to mostly gas-supported for the less-diffusive runs. For \HCR\ the median gas parcel is CR pressure-dominated out to roughly the virial radius. This broad spread in the pressure ratio is what affords our simulations such a rich CGM temperature structure, despite its uniformly diffuse gas density. This will have important implications for the metal columns explored in Section \ref{sec:cgm-metal-obs}.

\subsection{Cumulative Mass Profiles by Gas Temperature}
\stdFig{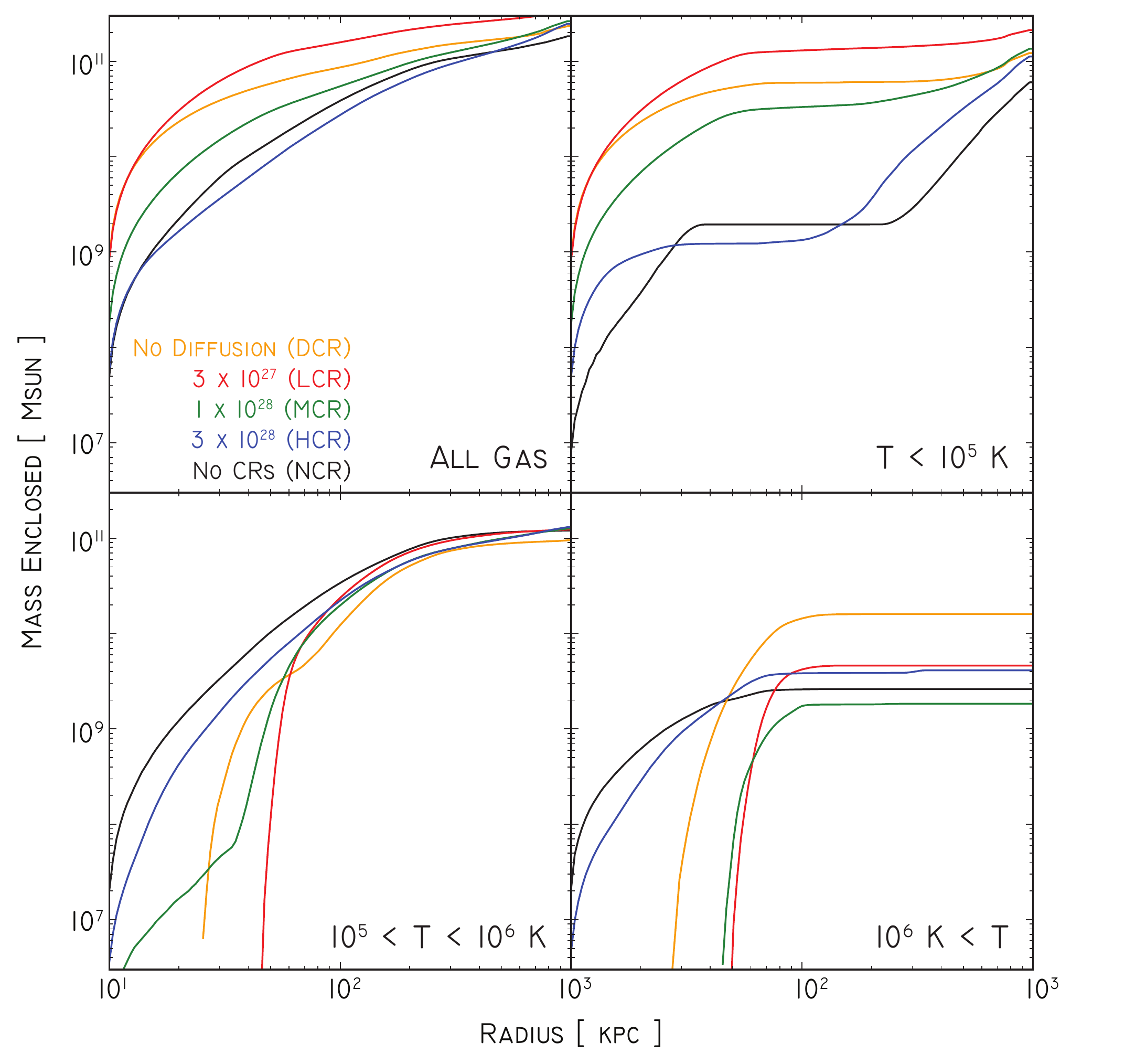}{Radial profiles of cumulative gas mass beyond 10 kpc from the galactic center at $z=0$. Clockwise from top-left, the panels show all gas, gas below $10^5$ K, gas where $10^5 < T < 10^6$ K and gas beyond $10^6$ K. Most notably, the mass of ``cold'' CGM gas -- below $10^5$ K -- within 200 kpc depends strongly on our CR diffusion model: raising the diffusion coefficient by a factor of ten, from $3 \times 10^{27}$ cm$^2$/s to $3 \times 10^{28}$ drops the cold gas mass of the CGM from $10^{11} M_\odot$ to just $10^9 M_\odot$, a factor of 100. }{fig:temp-mass}{.4}

We conclude this section by turning our attention to the total quantity of gas within our CGM, broken down by temperature component. As we've shown here and in SBH14, the shape, composition, density and temperature of the inner SF region varies substantially across each simulation's CR model. This complicates the task of delimiting where the disk-halo interface of our halo ends and where the ``CGM'' begins, especially since we've shown that cool temperatures and a wide range of densities exist to large radii in our CR-inclusive models. A conservative choice that avoids these issues is to simply exclude all gas within 10 kpc, regardless of physical state, which safely excludes the rotationally supported cold gas in all our runs, though it admittedly lobs off a good chunk of extra-planar material in the process. 

Figure \ref{fig:temp-mass} displays cumulative mass profiles of the gas beyond this 10 kpc demarcation, both for the entire CGM and also broken down by temperature. Within $\sim 50$ kpc, the total gas mass beyond 10 kpc varies by over a factor of 10 between the low-diffusion run ($\approx 10^{11} M_\odot$ at 50 kpc) to the most-diffusive run ($\approx 10^{10} M_\odot$). These mass disparities persist to the virial radius, though they shrink by a factor of $\sim 2$ in the process. Interestingly, the mass profiles of the two more-diffusive CR halos envelope the no-CR run's profile from low radius out to well beyond the virial radius: thus a realistic CR model does not seem to dramatically alter the total baryon content of the CGM.  We also note that the LCR (and DCR) runs have enhanced gas masses at the viral radius and somewhat beyond. We speculate that this occurs because the large CR pressures in these low-diffusion runs puffs out the gas distribution of infall halos, allowing its gas to be more easily stripped and incorporated into the main halo.

The most notable result of Figure \ref{fig:temp-mass} is the ``cold'' baryon content of our CGM, i.e. gas below $10^5$ K, which changes sharply across our diffusive CR runs. Between 10 and 50 kpc, the low-diffusion \LCR's halo contains over $10^{11} M_\odot$ of cool CGM --- a significant fraction of the halo's total baryon budget. In contrast, the \HCR\ simulation roughly parallels the \NCR\ run, with only $10^9 M_\odot$ within the same region. The \MCR\ run is unsurprisingly intermediate to these two cases, though closer to the less diffusive (and no-diffusion) run, with $M \sim 10^{10.5} M_\odot$. From this it seems a qualitative change in behavior for the CGM's structure occurs as the CR diffusion constant exceeds $\approx 10^{28}$ cm$^2$/s, suddenly more closely resembling the CGM of a halo devoid of cosmic rays, though more work is required to verify this, analyzing even more diffusive runs. The strong disparity in cold gas mass across runs persists out to beyond the virial radius, though the cumulative profiles begin to converge towards ``merely'' a factor of two difference at $\sim 1$ Mpc. At these large radii, all CR models contain substantially more cool gas than the non-CR run. We analyze the mass fraction of gas within these temperature ranges further in the discussion, with comparisons to analysis from recent CGM observations (see Section \ref{sec:mass}).

For our intermediate temperature cut of gas between $10^5$ and $10^6$ K, the mass within the virial radius is approximately constant across models. But once again, the picture within $\sim 50$ kpc is strongly model dependent. Now \HCR\ and \NCR\ feature substantially more intermediate temperature material between 10 and 50 kpc, while the less diffusive CR runs harbor progressively less gas. For \HCR, a hard barrier seems to exist at $\approx 55$ kpc, beyond which all gas is below $10^5$ K. These trends are the same for the highest temperature bracket of gas $>10^6$ K, though at large radii the balance never recovers, and the total mass is thus disparate across runs. Ignoring the unrealistic \DCR, the total ``hot'' CGM gas varies by $\pm 50\%$ across CR runs from the \NCR\ case, with no clear trend with diffusion coefficient, hovering around $\sim 10^{9.5} M_\odot$.

\section{Metal Line Observables}
\label{sec:cgm-metal-obs}

The analysis of Section \ref{sec:cgm-structure-comp} demonstrated that the CGM of our CR-inclusive runs were metal enriched to roughly $0.1 Z_{\odot}$ and featured a rich temperature structure, with far more cool, diffuse, CR-supported material than our run without cosmic rays. This result is especially intriguing given recent evidence from the COS-halos survey that diffuse, metal-enriched, relatively cool material may pervade the CGM of low-redshift $L \approx L^*$ galaxies \citep{Werk2014,Werk2013,Tumlinson2013,Tumlinson2011}. In this section, we analyze the metal content of our CGM and provide direct comparisons to metal columns inferred from the absorption features of background QSOs in the COS-Halos survey. To provide an accurate comparison, we depart from presenting results from our simulations at $z=0$ and instead analyze output files from $z = 0.2$, the typical redshift of the halos in the survey.

As described in Section \ref{sec:cgm-methodology}, our simulations explicitly track hydrogen and helium, computing the relative abundance of each ionization state (species) ``on-the-fly'', taking into account collisional ionization, photoionization, and recombination processes. For heavier elements however, we employ a simple bulk metallicity field, and use a \verb|Cloudy|-generated lookup table to compute its aggregate thermodynamic behavior. To generate column maps for these species, we thus reconstructed each ion's number density on a pixel-by-pixel basis during post-processing of the \verb|Enzo| data, utilizing \verb|yt| and additional lookup tables generated by \verb|Cloudy|. Details of this process are described in Corlies et. al. (2015), in prep, using methodology described in detail in \cite{Smith2011}. Briefly, the ion fraction lookup tables are functions of temperature and hydrogen number density, assuming solar metallicity and abundance ratios. The calculations depend on the ionizing metagalactic UV background, for which we employed the model of \cite{Haardt2001} with updates from 2005. This background is redshift dependent, and thus tuned to our epoch of inquiry at $z = 0.2$. With this ion abundance in hand, we compute a species' number density via
\begin{equation}
n_{X_i} = n_{\rm H}  \left( \frac{n_X}{n_{\rm H}} \right)  \left( \frac{n_{X_i}}{n_{X}} \right) \; .
\end{equation}
Here $n_{X_i}$ is the number density of the $i$th ionization state of species $X$ (HI at 13.6 eV, SiIV at 45.14 eV, CIII at 47.89 eV and OVI at 138.1 eV). The first ratio represents the elemental abundance relative to Hydrogen for which we assume $ ( n_{X} / n_{\rm H} ) = Z \cdot ( n_{X} / n_{\rm H} )_\odot$, i.e. the gas' explicitly tracked metallicity multiplied by the solar abundance ratio. The final ratio represents the ionization fraction interpolated from our \verb|Cloudy| tables.  We note that, in addition to the background radiation, ionizing sources in the disk will also contribute.  We do not include this effect, but as discussed in \citet{Hummels2013}, for the star formation rates we find at $z=0$ in our most realistic models, we expect it to only be important for $r < 25$ kpc. 

\stdFullFig{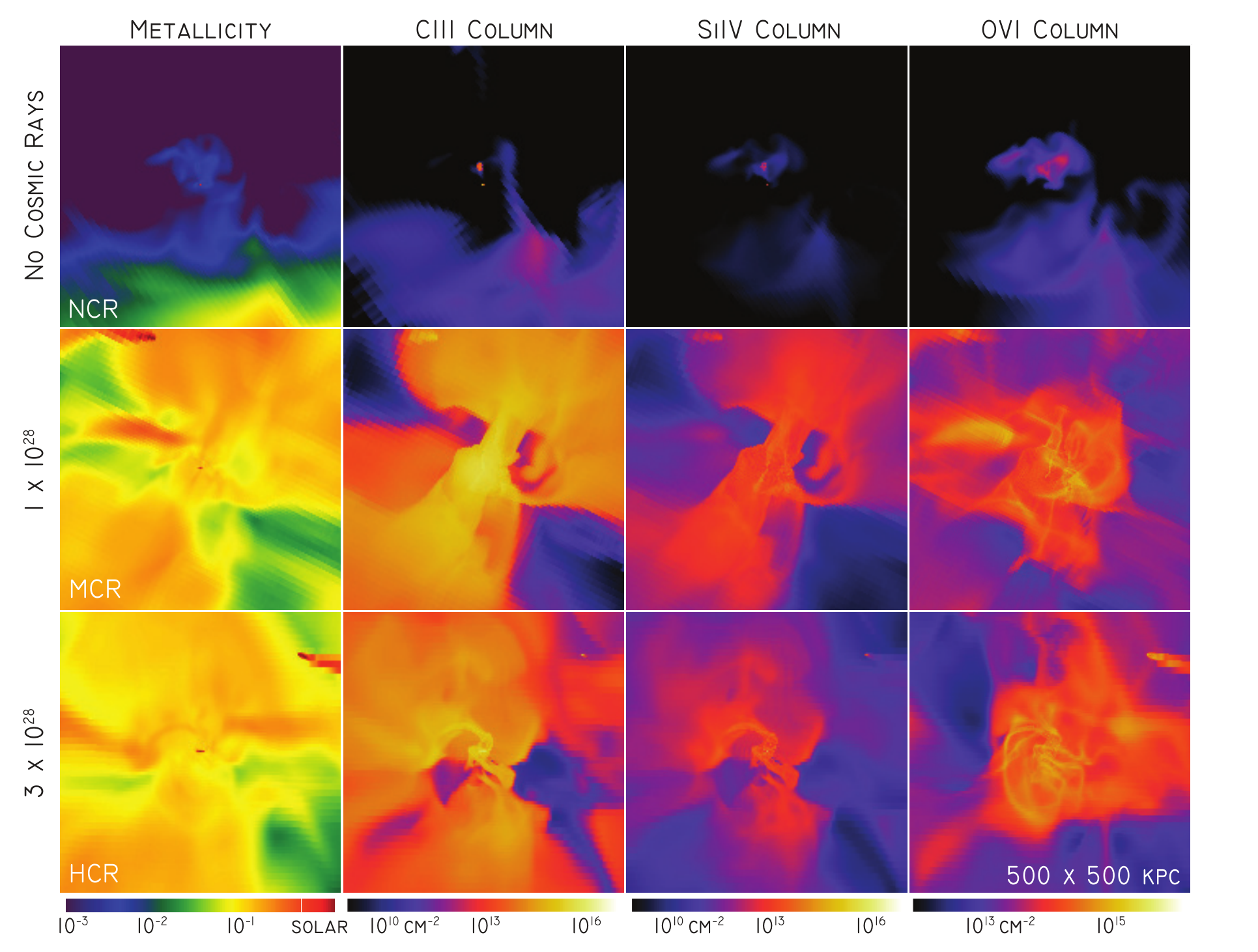}{From left to right: mass-weighted projections of metal density and column densities of CIII, SiIV and OVI ions. The CR-inclusive runs feature a much stronger presence of these ions, mostly due to the increased metallicity of those halos' CGM material.}{fig:chemistry}{.8}

\subsection{Column Density Maps}
\label{sec:cgm-metal-column-maps}

The middle two columns of Figure \ref{fig:chemistry} plot column maps for CIII and SiIV species with ionization energies of $47.89$ and $45.14$ eV, respectively, at $z=0.2$. These probe gas $\sim 10^{4.5}$ K. The disparity between the non-CR and CR-runs is enormous, with the former exhibiting almost no coverage at measurable column beyond the galactic center. For the CR-inclusive runs we chose to analyze in this section, both species show a covering fraction at $10^{13} {\rm cm}^{-2}$ of $\sim 1/2$ out to at least $r = 250$ kpc, though substantial gaps in coverage are evident in multiple quadrants at large radii. Examining the metallicity map, the coverage tracks strongly with the overall metallicity of the gas, though notable disagreements are evident, in particular among high-Z clumps towards the mid plane of both CR runs. Thus we find while including the CR-model leads to dramatically higher metal columns for ion species probing a range of ionization energies, the results are not strongly dependent on the choice of $\kappa_{\rm CR}$, at least for observationally motivated values $\kappa_{\rm CR} \sim 10^{28} \;{\rm cm^{2}/s}$.

The rightmost column of Figure \ref{fig:chemistry} plots OVI column across our simulations, again at $z=0.2$. OVI has an ionization threshold of $138.1$ eV, probing hotter gas at $\sim 10^{5.5}$ K, when collisionally ionized. From these plots we find the presence of OVI is dramatically higher in our CR-inclusive runs, which all feature a column $\sim 10^{14} \; {\rm cm}^{-2}$ with a significant covering fraction out to the limits of the plots at $r = 250$ kpc. In contrast, the non-CR run is all but devoid of OVI, with only an asymmetrical cloud of high metallicity gas in the right temperature regime in the southern half of the image (due to an absence of winds in the non-CR case).

\subsection{Radial Profiles}

\stdFullFig{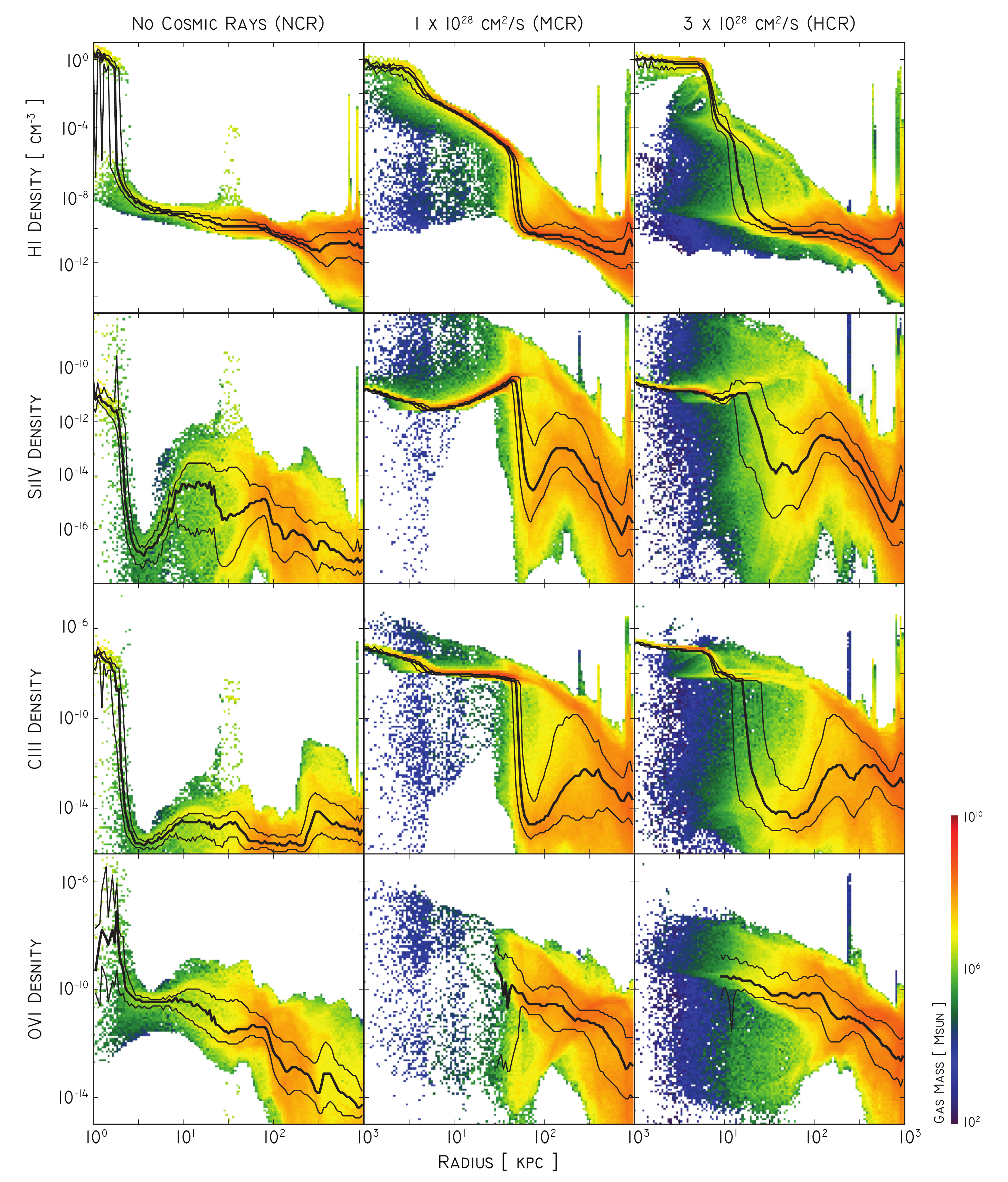}{Number density (in cm$^{-3}$) of important gas species, plotted against spherical radius as in Figure \ref{fig:profiles}. Colors denote total quantity of gas; lines show the median values and inner quartile range.}{fig:metal-profiles}{.5}

As in Section \ref{sec:cgm-radial-profiles}, we now turn our attention to the radial distribution of CGM metals. Figure \ref{fig:metal-profiles} shows radial profiles of number densities ($\rm cm^{-3}$) for HI, SiIV, CIII and OVI, along with median values at each radii and the inner quartile range. For brevity, we again plot only the non-CR \NCR\ run and our most successful CR simulations, \MCR\ and \HCR\ with $\kappa_{\rm CR} \in [ 1, 3] \times 10^{28} \; {\rm cm^{2}/s}$.

For HI, which probes gas with $T \lesssim 10^4$ K, the \NCR\ shows a small region of high density $\sim 1 \; {\rm cm^{-3}}$ in the disk, before dropping to $10^{-8} \; \rm cm^{-3}$ where it slowly falls off another two orders of magnitude by the virial radius. In contrast, \MCR\ features an almost constant falloff in column from within the disk out to intermediate radii $\sim 30$ kpc. Beyond this point, the column drops sharply, corresponding to a sharp rise in the median temperature (Figure \ref{fig:profiles}). \HCR\ is intermediate of these two extremes, with a clear falloff after the rotationally supported disk, but a respectable presence ($\sim 10^8 M_\odot$) of gas $\sim 10^{-4} \; {\rm cm^{-3}}$. 

For both SiIV and CIII, the non-CR run again exhibits substantially lower column in all but the central most SF region. For these ions, the \MCR\ run features a central plateau, corresponding to a region of predominantly $10^4$ K gas with $\sim 0.1 Z_\odot $ metallicity, before a steep drop-off at larger radii, where nevertheless the median number density is a factor of $\sim 100$ higher than the non-CR run. For the \HCR\ run, the behavior is similar, though with a larger spread in physical density at $r < 30$ kpc, which matches the corresponding larger spread in the temperature profile.

Finally for OVI, in the \NCR\ case, nowhere in the CGM does the median nor inner quartile range (IQR) of OVI density rise above a paltry $10^{-10}$ cm$^{-3}$. For the \MCR\ run, OVI is all but non-existent within the central $30$ kpc cavity of $\sim 10^4$ K gas, but it comes back strongly at larger impact parameter, where the density is nearly a factor of 100 higher than in \NCR. For the more diffusive \HCR\ run, the picture is similar to \MCR, though the median density is slightly higher at all radii, and the OVI presence penetrates deeper into the halo, unsurprising since higher temperature gas survives deeper into this run.

\stdFullFig{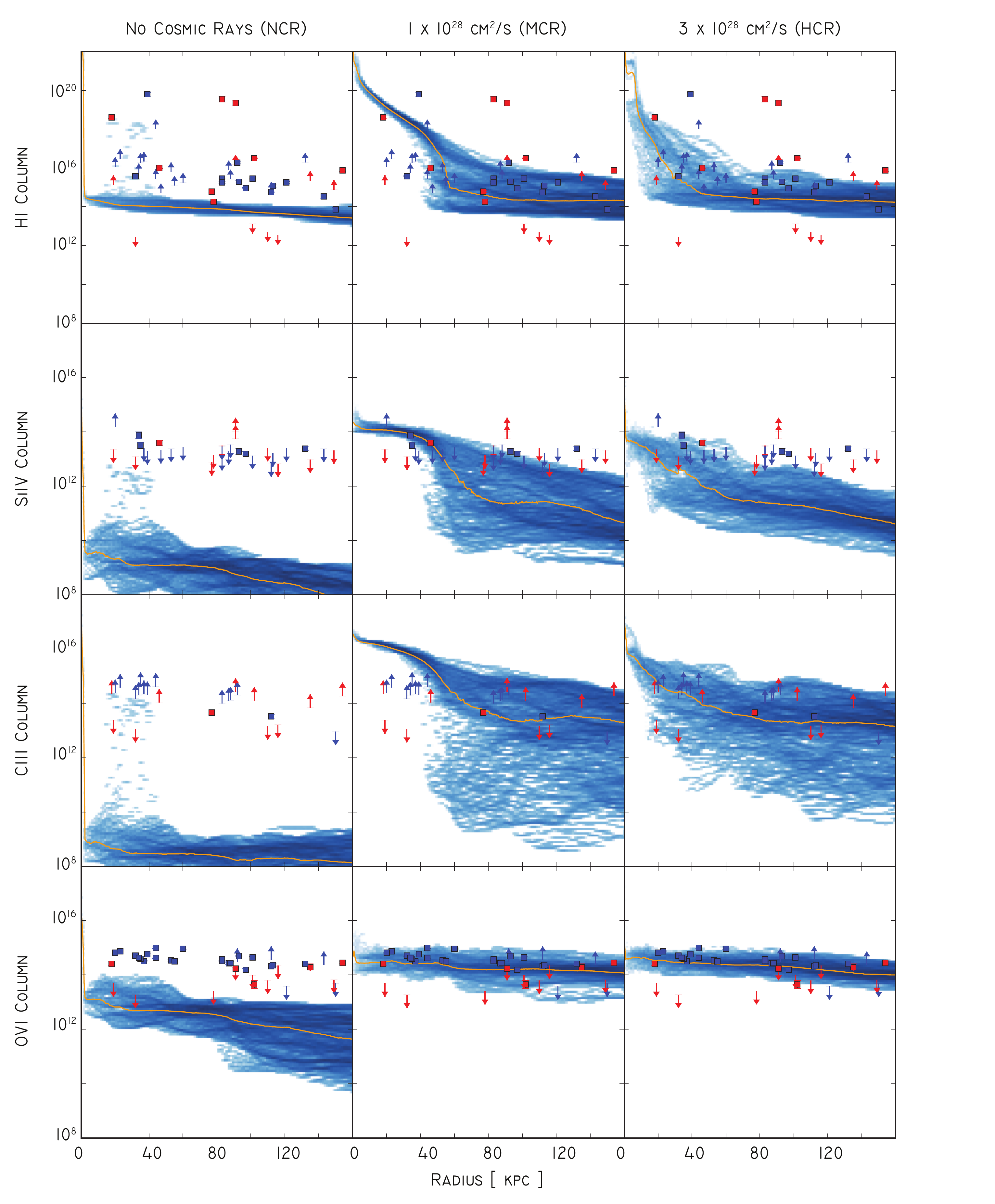}{A comparison of column density versus radius (``impact parameter'') between our simulations and quasar absorption line measurements from the COS-Halos Survey \citep{Werk2013}. The simulated profiles bin together all pixels from surface density maps generated from three orthogonal projections of the halo, with the median value at each radius shown as an orange line. Red and blue markers denote passive and star-forming galaxies from COS-halos \citep[demarcated at ${\rm sSFR} = 10^{-11} \; {\rm yr}^{-1}$ as in][]{Werk2013}. Squares show bounded measurements, whose errors are smaller than the markers shown here. Upward arrows represent saturated sight-lines, and thus lower limits, whereas downward arrows denote non-detections, and thus upper limits. Across ion species and impact parameters, the CR-inclusive runs (second and third colums) show better agreement with COS results.}{fig:cos-halos}{.45}

\subsection{Comparison to COS-Halos}
\label{sec:cgm-COS-halos}

Column maps similar to those in Figure \ref{fig:chemistry} allow us to describe the distribution of gas column as a function of impact parameter -- i.e., distance from the galactic center -- and thus compare directly to QSO measurements from the COS-Halos survey, using data presented in \cite{Werk2013} (hereafter W13). That work analyzed absorption line features in spectra of quasars behind 44 $L \approx L^*$ galaxies at $z \approx 0.2$. The sightlines probed CGM gas out to $r < 160$ kpc, revealing an unexpectedly high presence of far-UV absorbers, suggesting a massive CGM component at $T \approx 10^4 - 10^5$ K, perhaps accounting for a significant portion of an $L^*$ galaxy's baryon budget \citep{Werk2014}.


To provide a direct comparison, we took column maps out to $r = 160$ kpc along three orthogonal lines of sight\footnote{These column maps include gas along the line-of-sight out to the virial radius}, not necessarily aligned with the galaxy's disk. We then computed the distribution of column densities for HI, SiIV, CIII and OVI from these maps, as a function of radius. Figure \ref{fig:cos-halos} shows these profiles of column density, with the median values at each radii plotted as an orange line. Superimposed on the simulated data are measurements from COS-halos, broken down into red passive galaxies and blue star-forming galaxies \citep[the latter having sSFR $> 10^{-11}$, as in ][]{Werk2013}. Our simulations feature sSFR $\approx 4 \times 10^{-11}$, placing them comfortably in the star-forming category.  As before, the columns show the \NCR\ run and the two most successful CR runs: \MCR\ and \HCR.

Figures \ref{fig:basic-chem} and \ref{fig:metal-profiles} have already shown a far stronger presence of HI in the CR-inclusive runs. Thus it's no surprise that the HI column reaches higher values across all radii for the CR runs. Here \NCR's column distribution lies well below both the passive and SF detections at all radii (though above the passive non-detections). In contrast, the \MCR\ run's column distribution overlaps a majority of COS detections for both galaxy populations at impact parameters beyond $\sim 40$ kpc. Within 40 kpc, the distribution lies within the COS range, but with very little breadth, fails to overlap any detections. \HCR\ performs similarly at large impact parameter, though with a slightly tighter spread. But at small radii, it features a better overlap with COS detections, at least on the low end.

Likewise, the CR-inclusive runs perform noticeably better for SiIV. \NCR\ falls over 3 orders of magnitude lower than the COS detections and upper limits. The CR runs overlap nicely at low impact parameter, though both runs are distributed well below detections past $r \sim 50$ kpc (admittedly with many more upper limits in the COS data).

The story is similar for CIII, where again the non-CR run exhibits an enormous gap from the QSO measurements. Once again, \MCR\ and \HCR\ both provide good coverage: both runs span detections and  upper and lower limits at high impact parameter. At low impact, both runs are consistent with the saturated lines for SF systems, though \HCR\ features a broader range of columns, spanning down to the non-SF non-detections.

Finally, we look at OVI, where a clean separation between early and late-type systems emerges \citep{Tumlinson2011}. Once more \NCR\ produces columns roughly two orders of magnitude too low across the full span of impact parameters. The CR-inclusive runs perform much better, with a tight relation across impact parameter from $10^{14.5} \to 10^{14} \rm cm^{-2}$ from small to large radii. We explore the cause of this tight, flat relation in Section \ref{sec:cgm-discussion}. The median trend across these runs is nearly identical, though the \MCR\ run has a larger spread to low column at large impact. While our halo's span of column densities overlaps much better with the SF population than the early type systems (particularly regarding the non-detections) our median line lies below the typical SF system's column.

\section{Luminosity of Hadronic Losses}
\label{sec:cgm-hadron-luminosity}

A more direct measure of the relativistic component of the CGM comes from hadronic interactions between CR protons and the thermal gas. The CRs collide inelastically with gas particles in a process that yields charged and neutral pions, and the latter rapidly decay into GeV gamma-rays. This process is responsible for a majority of the diffuse high energy emission associated with the MW disk. Beyond the disk, the Fermi Gamma-Ray Space Telescope and predecessor instruments have identified a roughy isotropic gamma-ray background \citep{Abdo2010a}, generally considered the aggregation of numerous, unresolved extragalactic sources
\citep[e.g. starburst galaxies, gamma-ray bursts, large scale shocks. See, for example, ][]{Thompson2007,Keshet2003,Ackermann2012a}. Though often referred to as the ``diffuse extragalactic background'', a substantial fraction of this ``extragalactic'' background could in fact be due to the hadronic interactions of CR protons \citep{Feldmann2013}. At the very least, this observed background provides an upper limit of emission produced by our simulated CGM, and thus an important way to discriminate between our CR-diffusion models. In this section we compute the gamma-ray emission within our simulations at $z=0$, and compare to both the observed diffuse ``extragalactic'' background and a recent measurement of M31's gamma-ray halo. 

The emissivity from hadronic losses is related to the thermal and CR gasses by \citep{Ensslin2007,Jubelgas2008}
\begin{eqnarray}
\left.\frac{d \epsilon}{dt}\right)_{\rm had} 
	= -\frac{c \bar{\sigma}_{pp}}{2m_p} \rho \epsilon_{\rm CR}(q_{\rm thr})
\label{eq:hadronic-losses}
\end{eqnarray}
where the pion production cross section on average is $\bar{\sigma}_{pp} \approx 32$ mbarn \citep{Jubelgas2008}, $\rho$ is the thermal ISM's physical density and $\epsilon_{\rm CR}(q_{\rm thr})$ is the CR energy density of all CR protons above the energy threshold $q_{\rm thr} m_p c^2 = .78$ GeV. Our simulations do not track any measure of the CR gas's momentum distribution. For now we will assume the CR population of the CGM within our simulations is composed entirely of protons above this threshold. This simplification still affords us an upper limit on the luminosity due to hadronic losses. Appendix \ref{sec:cgm-spectrum} explores the uncertainty introduced by this assumption, to find a more detailed treatment lowers the emissivity by $10 - 70\%$, but not by an order of magnitude.

\stdFullFig{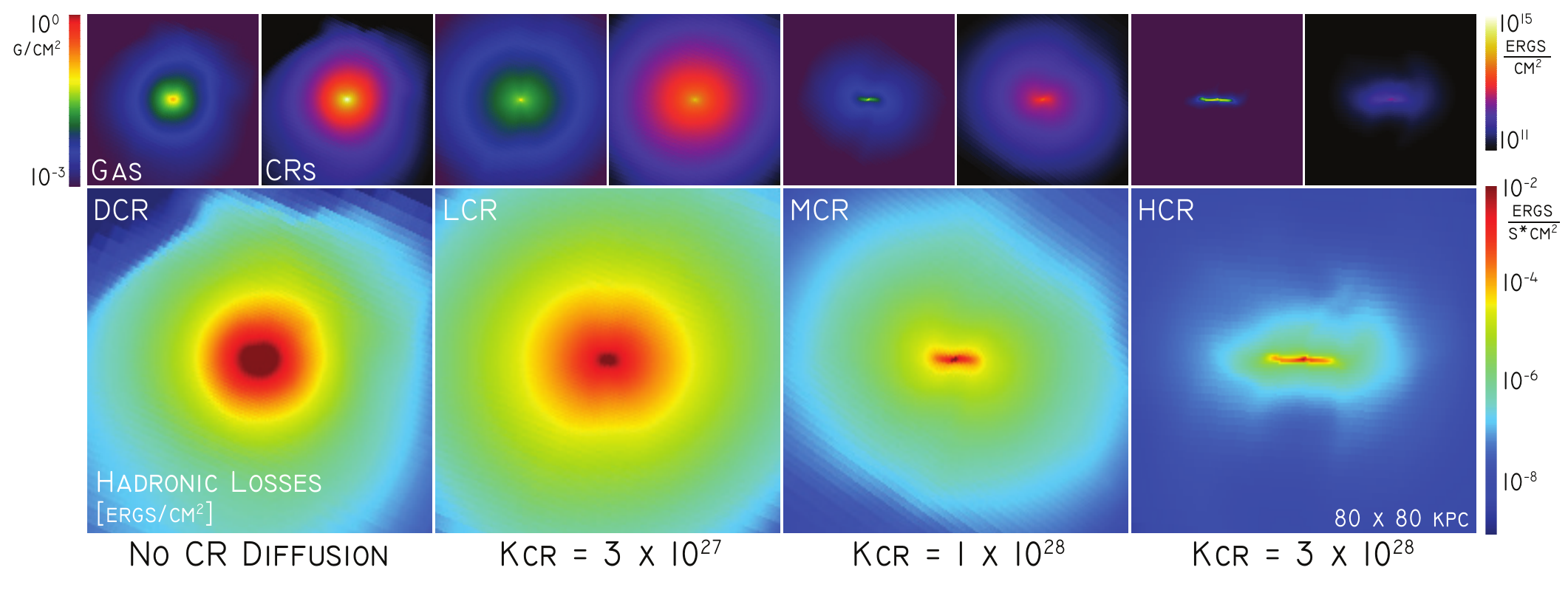}{Edge-on maps of $\gamma$-ray surface brightness due to hadronic losses across our four CR-inclusive simulations at $z=0$. The left-most run is devoid of CR diffusion, with the next three runs growing increasingly diffusive from left to right. Above each panel are thumbnails of the thermal gas and CR energy surface densities. Across runs, the central surface brightness exceeds $10^{-3}$ ergs/cm$^2\cdot s$. However, the falloff within the inner CGM is far more rapid for the more diffusive runs: a factor of 1000 lower for the $\kappa_{\rm CR} = 3 \times 10^{28}$ run, compared to the $3 \times 10^{27}$ run.}{fig:hadronic-losses}{.8}

We employed Equation \ref{eq:hadronic-losses} to compute the gamma-ray emissivity produced by CR protons throughout our simulation domain. We then used \verb|yt| to produce edge-on surface density maps of this emission, to provide a direct comparison to $\gamma$-ray observations. Figure \ref{fig:hadronic-losses} shows these mock observations across our four CR-inclusive simulations with various levels of CR diffusion. Across all our runs, the central surface brightness peaks beyond $10^{-3}\; {\rm ergs/cm^2/s}$, with the brightest regions tracing the densest gas coterminous with our simulations stellar disk/bulge, where both the thermal gas and CR energy densities are also at their maximal values. However, as we move tens of kpc from the galactic center, the falloff of this surface brightness varies dramatically across runs: roughly 30 kpc from the galactic center, halos most akin to a calorimeter (i.e. \DCR\ and \LCR) feature a surface brightness a factor of 1000 higher than the \HCR\ run, despite a mere factor of 10 difference in the diffusion coefficients. From Figure \ref{fig:hadronic-losses}'s thumbnails of thermal gas density and CR energy density, we see this difference is a compounded effect from the higher surface densities of both gas populations in the low-diffusion runs, where the CR gas manages to prop up an extended, spherical distribution of relatively high-density thermal plasma.

\stdFullFig{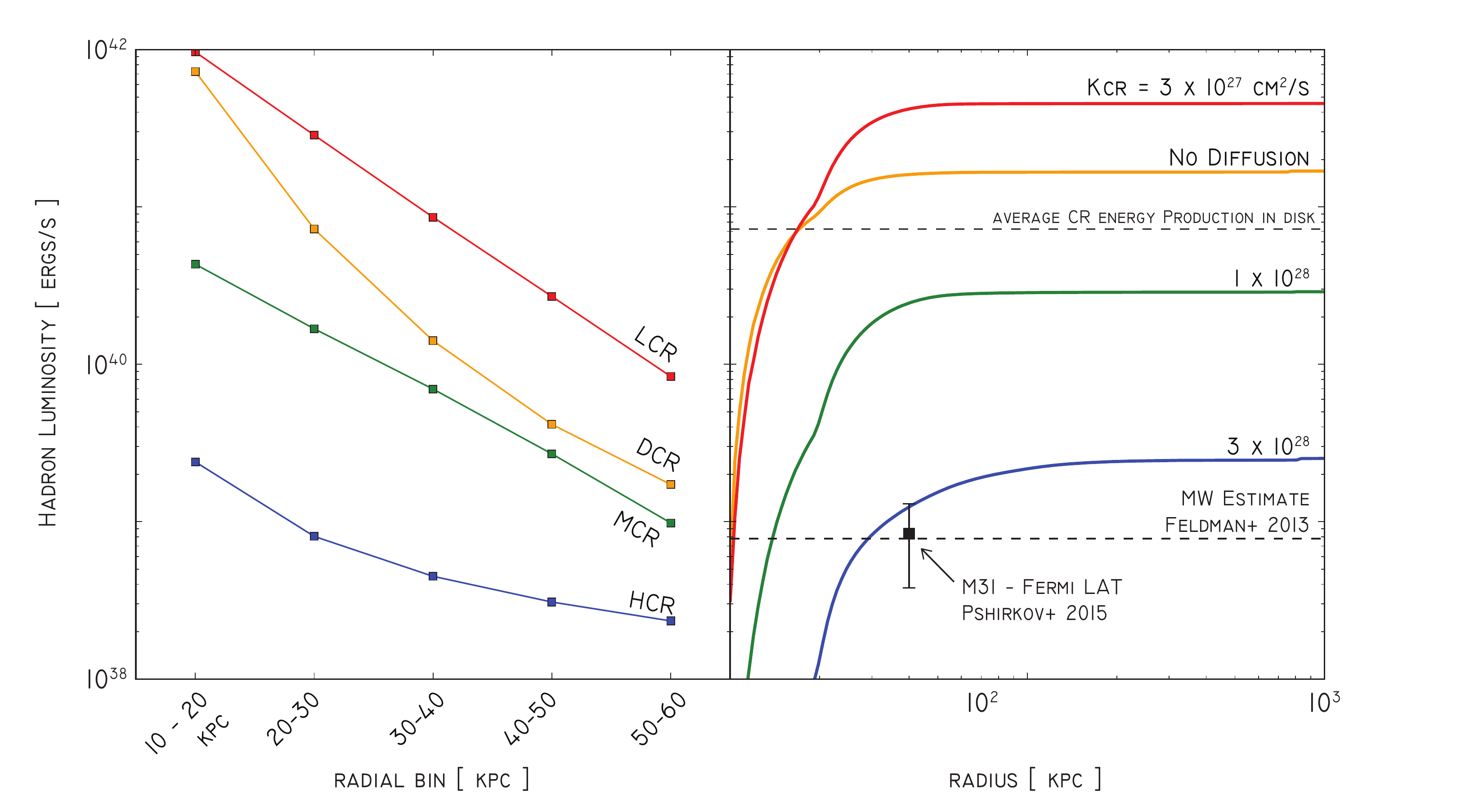}{Cumulative $\gamma$-ray luminosity from hadronic losses across our CR-inclusive simulations, as a function of radius from the galactic center at $z=0$. LEFT: Total luminosity emitted from radial shells about the galactic center of thickness 10 kpc. At all radii within the CGM, the diffusive CR runs separate strongly by the strength of the diffusion coefficient: a factor of 10 drop in the diffusion coefficient led to a brighter halo, by over a factor of 300 at $r \sim 20$ kpc, and nearly a factor of 30 at $r \sim 50$ kpc. RIGHT: Cumulative profile of luminosity as a function of radius, excluding a central disk of radius 20 kpc and height 10 kpc to safely avoid all emission from SF regions of the disk/bulge. The upper dashed line is the average rate of CR energy production by SF regions of the disk, and thus a gauge of where our choice to ignore hadronic losses in the simulation renders the model non-self-consistent. From this threshold, we find the \LCR\ and \DCR\ runs certainly fall into this category, while the \HCR\ run's emission is only $\approx 4\%$ this value. The situation is tenuous for the \MCR\ run. The lower dashed line shows an estimate for the MW's CGM from \citet{Feldmann2013}. Also shown is a recent measurement of diffuse halo emission around M31 from 5 years of Fermi LAT data  \citep{Pshirkov2015}. Across the diffusive runs, a factor of three increase in the diffusion coefficient causes this total luminosity to drop by over an order of magnitude. This suggests matching the CGM's $\gamma$-ray emission provides a tight constraint on models of the thermal gas/CR interaction.}{fig:hadron-luminosity}{0.5}

We next compute the total luminosity of $\gamma$-ray emission due to hadronic losses within the inner CGM of our simulated galaxies, by summing the emission described by Equation \ref{eq:hadronic-losses} pixel by pixel. Figure \ref{fig:hadron-luminosity} shows this result across our 4 CR simulations. The left panel shows sums computed within radial ``shells'' of thickness 10 kpc. Immediately evident is a strong separation across our three diffusive CR models, where at $r \sim 20$ kpc, a factor of 3 increase in the diffusion coefficient leads to a rise in the local luminosity by over a factor of 10; between the most and least diffusive run, the inner CGM emission drops by a factor of over 300. The contribution to the luminosity from progressively farther our shells drops substantially, despite the growing volume of these regions, as both the CR energy density and gas density drop precipitously across all runs. However, the slope of the falloff is markedly shallower in \HCR, and so the gap in emissions between runs closes to a factor of $\sim 30$. Interestingly, \DCR\ (which SB14 found incapable of driving strong winds from SF regions) shows a very different radial behavior, matching the low-diffusion run close to the disk, but paralleling the \MCR\ run at larger radii.

With the total gamma emission in hand, we can investigate whether the choice to ignore hadronic losses during our simulations is truly self-consistent. To judge this, consider the average total production of CR energy within the star forming disk. As described in Section \ref{sec:mass}, the total stellar mass of our simulations is $\approx 7 \times 10^{10} M_\odot$ across our runs (somewhat higher for \DCR\ and lower for \LCR, though these runs are least favored by our analysis). The total production rate of cosmic rays in energy is thus given by $7 \times 10^{10} M_\odot \times c^2 \epsilon_{\rm SN} \times f_{\rm CR} / {13.8 \rm Gyr} = 8.6 \times 10^{40}$ erg/s. From Figure \ref{fig:hadron-luminosity} we find this estimate renders the \DCR\ and \LCR\ runs non-self-consistent, since their hadronic loss output exceeds this value, by a wide margin. In contrast, the high diffusion \HCR\ run has a hadronic emissions at $4\%$ of this value. For the intermediate case of \MCR, the situation is more uncertain. Here hadronic emissions are likely important to the dynamics, and could possibly degrade the gamma-ray emission in this model to an extent that placed it closer to the observational constraints.

These results suggest the gamma-ray emission in the galactic halo is strongly sensitive to the details of CR propagation in the galactic halo, and thus comparisons to observed diffuse gamma-ray emission can constrain our models. Recently, \cite{Feldmann2013} estimated the contribution of CR protons to gamma-ray emission from the MW's CGM, finding emission related to CR proton interactions with the thermal CGM could contribute $3 - 10\%$ to the isotropic background observed by Fermi. This past year, \cite{Pshirkov2015} used over 5 years of Fermi LAT data to constrain this luminosity for M31 within 40 kpc of the galactic center, excluding the disk, to $8.4 \pm 4.6 \times 10^{38}$ ergs/$s^{-1}$. The right-hand panel of Figure \ref{fig:hadron-luminosity} compares these results to our simulated CGM. The plot shows cumulative luminosity profiles including only gas beyond a central disk of radius 20 kpc and height 10 kpc from the galactic center, in order to avoid contributions from the central SF disk/bulge. From this we find our most-diffusive \HCR\ run, with $\kappa_{\rm cr} = 3 \times 10^{28} \; {\rm cm^{2}/s}$, comes closest to matching these estimates. While these luminosities depend on knowledge of the CR momentum spectrum, which we do not explicitly track in our simulations, Appendix A shows this uncertainty can only change the results by a factor of 2, well within the decade of separation in luminosity between our CR models.   Finally, we point out that the gamma ray fluxes computed can be viewed as lower limits, as there may also be a contribution from inverse Compton scattering off of an electron CR component.  In this paper we did not model the CR electrons because of their short cooling time and negligible impact on the dynamics.  Their contribution to the gamma-ray flux near the disk is generally thought to be small \citep[e.g.,][]{Orlando2015}, although it may be larger in the halo.

\section{Discussion}
\label{sec:cgm-discussion}

\subsection{Constraining the CR Diffusion Coefficient within the CGM}
\label{sec:cgm-diffusion-coefficient}

A key result of the present work is the central role of the cosmic ray proton fluid's diffusion parameter, $\kappa_{\rm CR}$, which provides a first-pass at capturing the exchange of CRs between gas fluid parcels, and whose value can substantially alter the behavior of galactic-scale flows. SB14 found simulations of an idealized gas disk galaxy within a $10^{12} M_\odot$ halo with $\kappa_{\rm CR} \in [ .3 , 1 , 3 ] \times 10^{28}$ cm$^2$/s were all capable of launching winds from the forming system with mass-loading factors on order unity, with a clear trend of higher mass-loading factors and lower SFRs for lower values of $\kappa_{\rm CR}$. SBH14 then applied the model in a cosmological setting with the same simulations analyzed in the present work, and found the simulation employing the highest of these values (referred to as the \HCR\ run here) produced the least-peaked rotation curve and the thinnest, most extended stellar and gaseous disks at low redshift. This value of $\kappa_{\rm CR} = 3 \times 10^{28}$ cm$^2$/s is in line with a range of observational measurements of the diffusion parameter for the central star-forming region of galaxies \citep{Strong1998,Ptuskin2006, Ackermann2012,Tabatabaei2013}. In the present work we extended the analysis into the diffuse CGM of our $10^{12} M_\odot$ halo, where the true value of diffusion coefficient and indeed the validity of treating the process with a homogenous, scalar coefficient is even less certain.

Our simulations have shown the choice of diffusion coefficient sets the size of a central, CR-dominated spheroidal region within the CGM, where the gas temperature is uniformly low, at the $10^{4}$ K floor of our runs. The size of this region corresponds to the spatial scale $L$ at which the CR diffusion timescale, $t_{\rm CR, diff} = L^2 / \kappa_{\rm CR}$ exceeds the dynamical time, corresponding to 10's of kpc for the diffusion parameters explored here. This strongly affects the observational signatures of our CGM, with less diffusive runs featuring enhanced HI columns and an absence of OVI within this region. Thus the presence of UV absorbers at low impact parameter can discriminate between diffusion models in the inner CGM; indeed the match to COS-Halos seen in Figure \ref{fig:cos-halos} already suggests values of $\kappa_{\rm CR} \lesssim 10^{28} \; {\rm cm^2/s}$ produce an excess of HI at low impact parameter. In section \ref{sec:cgm-hadron-luminosity} we found the total hadronic emission within this region grows dynamically important in runs with lower diffusion coefficients, casting doubt on whether or not a more sophisticated model would produce such a bubble.

Exterior to this central region, the metallicity, temperature and density structure of the CR-infused CGM looks roughly equivalent across diffusion models. All our CR-inclusive models successfully populate the CGM with enriched material, raising the metallicity to $\approx 0.1 Z_\odot$ across runs, as compared to a run devoid of cosmic rays. At these length scales, diffusion has become of secondary importance to the dynamics, with bulk transport and radiative heating and cooling driving the dynamics, and thus it is unsurprising the outer CGM's appearance is roughly independent of the CR diffusion coefficient.

Our explicit treatment of the high-energy ISM also permitted a prediction for the CGM's diffuse gamma-ray emission, which we found to be quite sensitive to the choice of diffusion coefficient. Figure \ref{fig:hadron-luminosity} shows a factor of 3 increase in the diffusion coefficient caused the gamma-ray luminosity to drop by a factor of 10, regardless of where in the CGM you choose to collect photons. \HCR, the $\kappa_{\rm CR} = 3 \times 10^{28}$ cm$^2$/s run, provided the closest match to preliminary observations of the diffuse gamma-ray emission from Andromeda's CGM. In addition, this run's value fell below the level of the diffuse ``extra-galactic'' emission measured by Fermi, though in excess of estimates from the MW halo's contribution. The direct comparison here is tricky, since our model fails to track the distribution of CR protons in momentum space, and thus we may be including lower energy CRs that do not participate in the production of GeV gamma-rays (though Appendix A suggests this will only alter results by a factor $\sim 1/2$). In addition, the \LCR\ and \DCR\ runs are inconsistent with observations due to a high rate of hadron losses that ought to become dynamically important. For the \MCR\ run, these losses may likewise alter the dynamics, pulling the emission more in line with observations. For all these reasons, Figure \ref{fig:hadron-luminosity} should be taken as an upper-limit for each model. Nevertheless, from the strong separation it appears the most diffusive model provides a superior match to preliminary observations and more diffusive halos could perhaps do even better. The mean-free path of CRs is likely to rise in the galactic halo, compared to the disks of galaxies, and thus it seems likely the CR proton fluid grows more diffusive at large radii. Thus models with a relatively low $\kappa_{\rm CR}$ within star forming regions may still be valid, provided the underlying magnetic field structure changes substantially in the CGM. Since the majority of gamma-ray emission occurs within $\sim 40$ kpc of the galactic center, this transition would likely need to occur within the disk-halo interface.

\subsection{Role of CR Pressure Support in the CGM}
\label{sec:cgm-cr-support}

Beyond simply driving winds (and thus metals) into the halo, the CR fluid's pressure support is of crucial importance to the structure of the CGM at all radii. As discussed above,  in the inner regions of the CGM, corresponding to $r$ such that $r^2/\kappa_{\rm CR} < t_{\rm dyn}$, CR pressure overwhelmingly dominates, producing a region rich in HI and $10^4$ K gas. At larger radii, CRs provide an intermediate level of pressure support that allows broad swaths of diffuse gas to cool, spanning a wide range of temperature down to $\sim10^{4.5}$K. Only a small fraction much exceeds $10^6$ K. 

Thus the presence of CR pressure support in the galactic halo has decoupled the diffuse medium's temperature from the halo's virial temperature without forfeiting hydrostatic balance. The broad temperature range of our CGM, coupled with the robust flow of metals out beyond SF regions of the disk provided a good match to COS-halos data for a set of species with disparate ionization energies. This result separates the CR-inclusive model considered here from ad-hoc wind models, which in particular have struggled to reproduce the observed OVI column \cite[e.g.][]{Ford2015,Hummels2013}.

\subsection{Role of Cooler Diffuse Gas in Observed Ion Columns}
\label{sec:cgm-cool-gas}
How important to observed ion columns is the relatively cooler, diffuse CGM material discussed in Section \ref{sec:cgm-cr-support}? From Section \ref{sec:cgm-structure-comp}, it's clear the greater success of the CR-inclusive runs strongly depends upon metal enrichment, as winds from the disk enrich large swaths of the CGM to roughly $1/10$th $Z_\odot$ in the CR runs but not in \NCR. But the presence of metals is perhaps not the whole story. Here we briefly explore the role of colder diffuse CGM material in providing a superior match to the COS-halos data than enriched winds alone. 

To explore this issue, we replicated the analysis of Section \ref{sec:cgm-metal-obs} using data from our simulations but artificially fixing the metallicity of the CGM to a uniform value of $1/10$th $Z_\odot$. This eliminates any features in the ion column maps due to variations in metallicity and ``upgrades'' the \NCR\ run substantially by providing the ion densities necessary to match observables. The three runs feature comparable CGM gas density profiles, so this experiment isolates disparities in gas temperature caused by the presence of CRs. If this enriched material driven by winds to the CGM were the whole story (i.e. factor-of-three differences in the gas temperature were not important) then the \NCR\ run should now produce an equally good match to the COS-halos data.

\stdFullFig{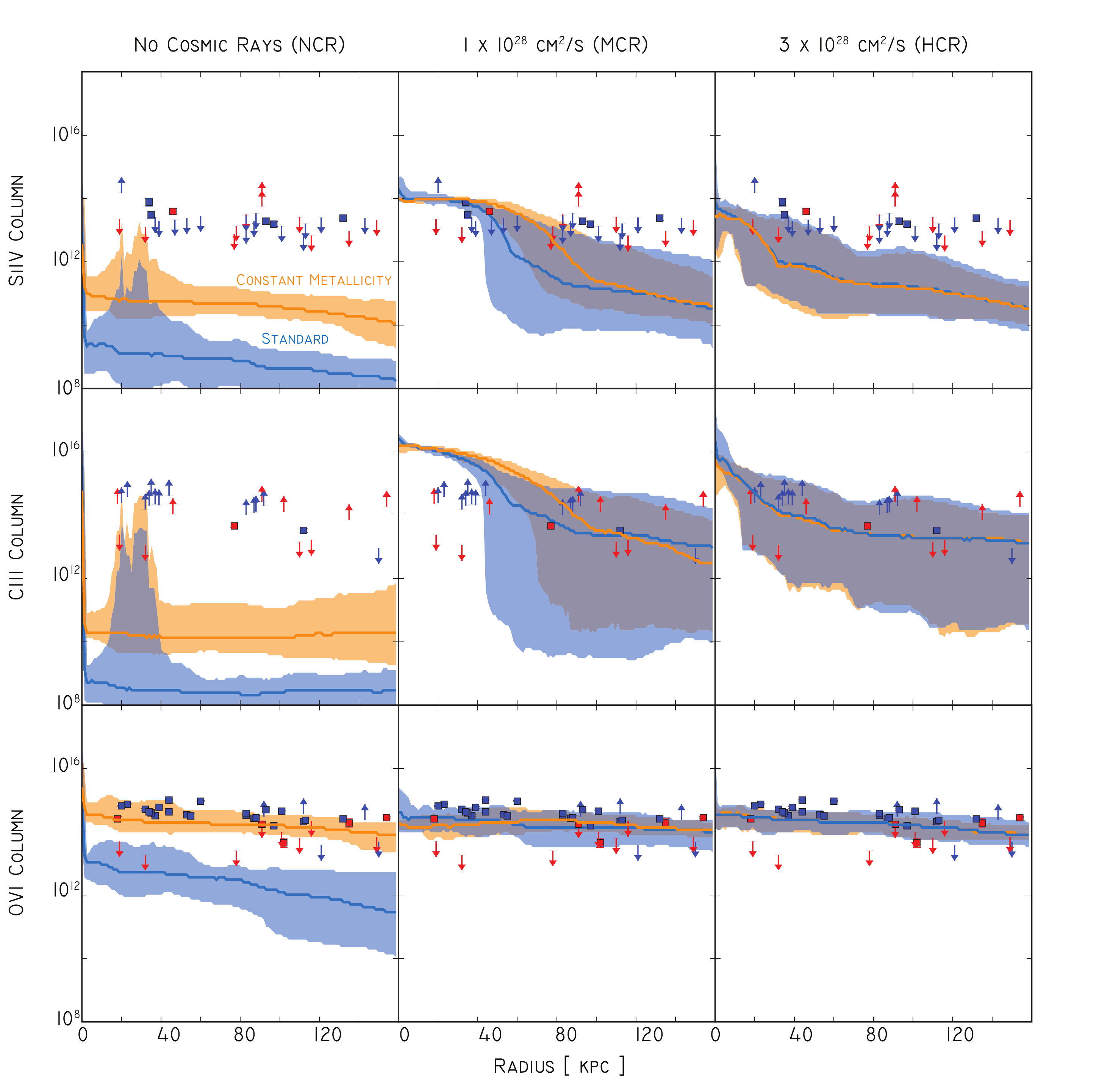}{A measure of ion column density versus impact parameter across three of our simulations, computed as in Figure \ref{fig:cos-halos}, with quasar absorption line measurements from the COS-Halos Survey again over-plotted. In addition to the standard analysis (blue), we plot an identical analysis where metallicity throughout the CGM was artificially fixed to $0.1Z_\odot$ (orange). For clarity, we omit the full distribution of column an instead show the 95\% range (shaded regions) and the median value (lines) versus impact parameter. The COS-halos data is plotted in an identical fashion to Figure \ref{fig:cos-halos}. Although artificially enhancing the metallicity of the \NCR\ run (left column) boosts the metal columns across impact parameter, this enhanced metallicity (together with a comparable density profile to the \MCR\ and \HCR\ runs) is not enough to produce a match to the observed lines for SiIV and CIII. This shows the presence of cooler, diffuse CGM material (as in \MCR\ and \HCR), via CR pressure support, is a necessary component in matching a broad array of ion columns. }{fig:COS-metallicity}{.5}

Figure \ref{fig:COS-metallicity} shows the result of this experiment. As in Figure \ref{fig:cos-halos}, we produce metal ion column maps along three random sight-lines and computed the distribution of ion-column as a function of impact parameter (radius from the galactic center), now for both the standard analyses and an analysis at a fixed metallicity of $0.1 Z_\odot$. For the \NCR\ run (first column) devoid of CR physics, simply enhancing the metallicity was not enough to produce good matches for the SiIV and CIII columns; while both rise by an order of magnitude across the whole range of impact parameters, both still undershoot the COS-halos columns by an order of magnitude. For OVI, the story is different: the artificially enhanced metallicity has allowed the \NCR\ run to match observations. For the \MCR\ run, the fixed metallicity has decreased the variability of column along sight-lines at low impact parameter, now missing some lower limits from the star-forming COS halos. But otherwise, both the \MCR\ and \HCR\ results are mostly unaffected by the fixed metallicity. 

These results demonstrate that metal enriched winds from the disk plane are not enough to produce a realistic CGM capable of matching observables. The addition of cooler, diffuse material, with pressure support from cosmic rays, is necessary in these models to match a broad array of ion columns.

\subsection{Baryonic Mass within the CGM}
\label{sec:mass}

There exists a long-running discrepancy between the cosmic mean baryon fraction, $\approx 17\%$ of matter, and the readily-observed baryonic mass components within galaxies, such as stars and the ISM. While this ``missing'' material may have been expelled from halos via sufficiently mass-loaded winds, observations thus far of the low-redshift intergalactic medium (IGM) fail to account for the total balance of baryons \citep[e.g.][more recently]{Cen1999,Prochaska2011}. The difficult to observe, diffuse, gaseous CGM has been invoked as a potential line item capable of repairing a galaxy's baryonic balance sheet. Classic analytic arguments \cite[e.g.][]{White1978} positing a halo chock-full of diffuse, virialized material, $\sim 10^{6}$ K for a MW-sized system, suggested these baryons would exist in the form of hot X-ray absorbing gas. However, recent observations of the MW and surveys of $L \sim L^*$ galaxies have cast doubt on this assertion \citep{Benson2000,Miller2013,Anderson2013}, although see \cite{Gupta2012}. Studies of UV absorbers at low-redshift, culminating in the COS-halos survey, now suggest diffuse, hard to observe material could indeed still account for a significant portion of a galaxy's baryons, though material at a substantially lower temperature: perhaps nearly half of a galaxy's ``expected'' baryons exist in diffuse, $T < 10^5$ K material  \citep{Werk2014,Tumlinson2013}; and a warm-hot $10^{5} - 10^{7}$ K mode potentially harbors a similar quantity of mass \citep{Peeples2014,Tumlinson2011}.

\stdFullFig{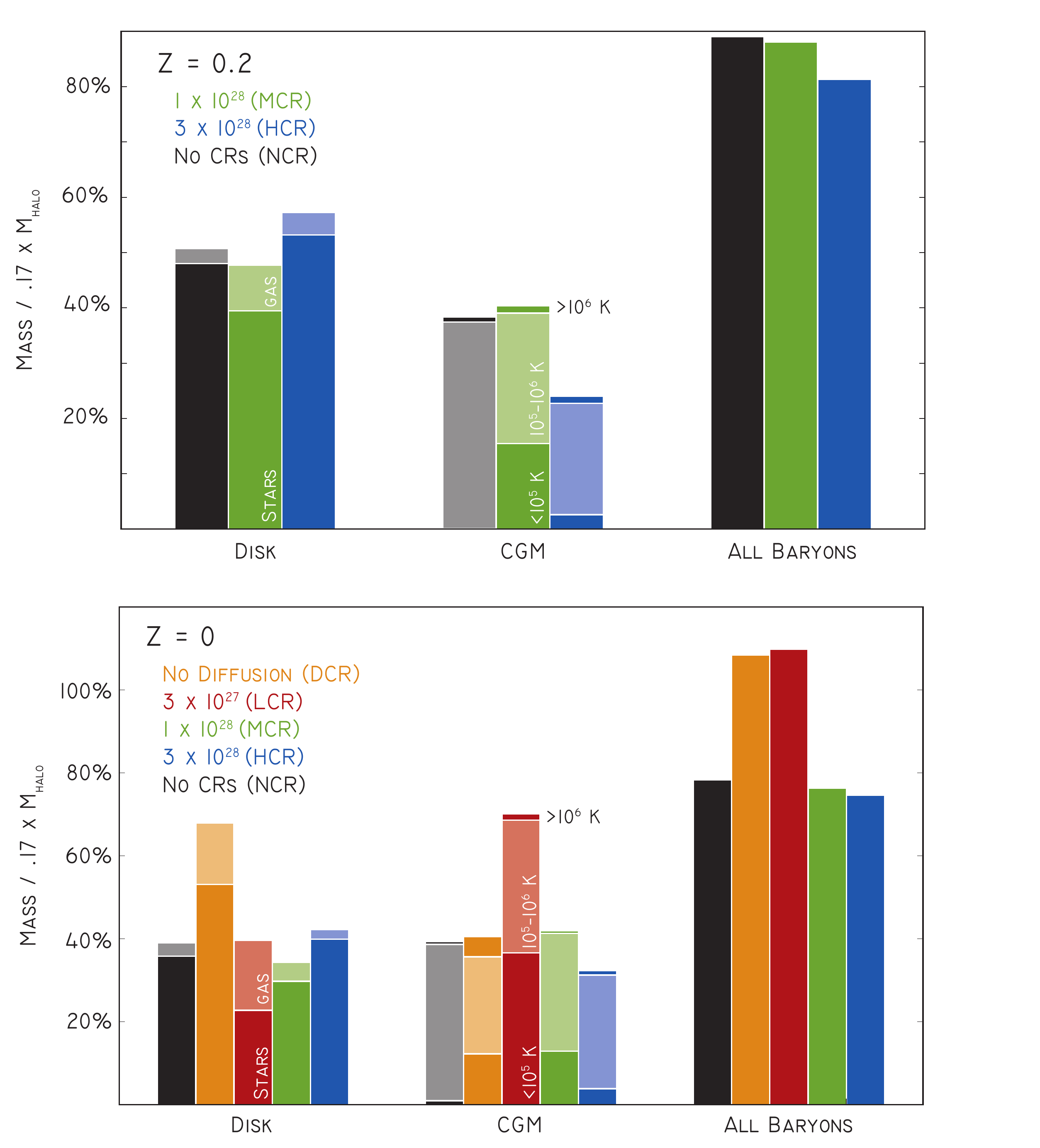}{Mass components of our simulated halo, normalized to the ``expected'' baryon mass of $M_{\rm Halo}(300 \; {\rm kpc}) \times \Omega_b/\Omega_m$. Here the ``disk'' includes all gas within a cylindrical radius $R < 20$ kpc from the galactic center and a height $z < 10$ kpc from the galactic mid-plane. The CGM is gas beyond this disk but within $300$ kpc. TOP: The \NCR, no-CR run, compared with our most successful CR diffusion models, \MCR\ and \HCR, at $z=0.2$ for a direct comparison to COS-halos. BOTTOM: The same quantities at $z=0$, for all runs.}{fig:mass-frac}{.5}

Figure \ref{fig:mass-frac} shows the total balance of baryons within our simulations, for a ``disk'' component, defined as stars and all gas within a cylindrical radius $R < 20$ kpc and height $z < 10$ kpc of the galactic center, and a CGM component, comprising all remaining gas within $300$ kpc. These values have been normalized to the total ``expected'' baryon mass, defined as the DM mass within 300 kpc multiplied by the cosmic fraction, $\Omega_{b} / \Omega_M = .17$. 

The top panel shows results for the \NCR\ run against our two most successful CR-inclusive runs, \MCR\ and \HCR. The mass within the disk of all three runs is $\approx50\%$, which is over double the $\approx 15 - 25\%$ found observationally \citep{Behroozi2010}. As noted in SBH14, while the CR model persistently drives galaxy-scale winds from the disk, promoting SF in a thin extended disk, it fails to fully disrupt the runaway star formation at the galactic center, due either to a lack of sufficient resolution or the absence of other feedback processes relevant on small scales (e.g. turbulence, photon pressure or a more sophisticated treatment of SN and stellar winds). This problem becomes more accute for the \HCR\ run, which actually forms more stars than the \NCR\ case. Within the CGM however, this situation is reversed: while \NCR\ and the moderately-diffusive \MCR\ feature $\approx40\%$ of ``available'' baryons in their CGM, the \HCR\ run has managed to drive a substantial fraction of baryons from the halo entirely. Indeed this run features an aggregate baryon fraction of only $80\%$ at this epoch within 300 kpc, in comparison to the other runs' $90\%$. Between the CR and non-CR runs however, the most profound change involves the \emph{temperature} of the gas. With the advent of CR-supported diffuse material in the halo, as discussed in Section \ref{sec:cgm-cr-support}, the CGM moves from being entirely warm-hot gas above $10^6$ K to harboring a substantial fraction of mass in a cool, $T<10^5$ K phase: $\approx 15\%$ for \MCR\ and a more modest few percent for the \HCR\ run.

This picture persists to low redshift, where the disk's contribution to the baryon balance shrinks to less than 40\%, while the CGM fraction holds steady. Here we show all five simulations, where all but the most-diffusive \HCR\ run harbor a substantial fraction of cold CGM material, most notably the $3 \times 10^{27}$ cm$^2$/s \LCR\ run. Both \LCR\ and the diffusionless \DCR\ simulation are outliers in either disk or CGM mass: \DCR\ formed substantially more stars than any other run in a central, dense bulge. Meanwhile \LCR's modestly extended disk was surrounded by a large reservoir of cool, CR-supported gas, accounting for roughly $40\%$ of the baryon budget. Both runs feature substantially more mass in the entire halo, with $\approx 100\%$ of baryons accounted for. However, the results of Section \ref{sec:cgm-hadron-luminosity} suggest these scenarios predict a far too bright gamma-ray halo for the MW, in addition to involving diffusion coefficients inexplicably lower than what's observed in the solar neighborhood.

\subsection{Missing Physics}
\label{sec:cgm-limitiations}
Our simple cosmic ray model affords an unprecedented look at the the dynamical impact of high energy particles in high resolution global galaxy simulations. However, this two-fluid, isotropic diffusion approach is only a crude approach to the complex interaction between CRs, magnetic fields and the thermal gas. We provide a brief summary of the most important missing physics in the context of CRs.

Our runs are devoid of explicit magnetic fields. A proper treatment of MHD would require an inaccessible level of resolution for today's generation of cosmological zoom simulations. The validity of our two-fluid model rests on the assumption that twisted, stochastic magnetic fields are present, to some degree, throughout our simulation domain. In the galactic plane, there is a rough equipartition in energy between such a stochastic field and a more large-scale, coherent component following spiral structure \citep{Beck2013}. The CGM's magnetic field structure however is highly uncertain, and earlier work has sometimes modeled its structure as a coherent, radial field upon which CRs may coherently stream \citep[e.g.][]{Breitschwerdt1991}. It is unclear how streaming would change this picture, though there is evidence streaming CRs can deliver heat and momentum to the thermal ISM, driving winds and providing pressure support via MHD waves \citep{Zirakashvili1996,Everett2008,Uhlig2012,Dorfi2013}. Evidence from abundance ratios of CRs in the solar neighborhood suggest that a fraction of CRs having returned from excursions into the Galaxy's halo have a lifetime of $\sim 20$ years in the galaxy \citep{Shapiro1970,Kulsrud2005}, which suggests a degree of confinement for CGM CRs.

Our simulations account for adiabatic transfers of energy between the CR and thermal fluids and the acceleration of CRs in star forming regions, but otherwise are devoid of cosmic ray gain and loss processes. A more detailed treatment would include a source term for CRs formed during AGN activity at the galactic center and CRs accelerated at cosmological shock fronts. Several important loss processes have also been ignored here, including Coulombic and hadronic losses of CRs as they interact directly with the thermal gas. The former process is of more importance for low-energy CRs and in regions of higher density, and thus we expect our CR spectrum to be quite hardened by the time the fluid parcel has reached the CGM, where the surviving highly-relativistic rays travel through a far more rarefied medium. The latter process was discussed in Section \ref{sec:cgm-hadron-luminosity}, where we noted that the lower diffusion runs were not self-consistent due to a luminosity via these hadronic losses in excess of the average rate of CR energy production within the disk. Thus hadronic losses may measurably deplete the CR population, but likely do not kill their strong presence in the halo. CRs also excite MHD waves which also transfer heat to the thermal ISM \citep[e.g.][]{Kulsrud1969,Skilling1975,Cesarsky1980}, a process that may become increasingly important on the longer timescales of these dynamics relative to in the galactic plane. Given the purported dynamical impact of CRs in the present work, this suggests a detailed understanding of the CGM's plasma behavior remains critical to characterizing its composition. Our model also fails to resolve the distribution of CRs in momentum space, which we explore in the context of catastrophic losses in Appendix A.

\stdFig{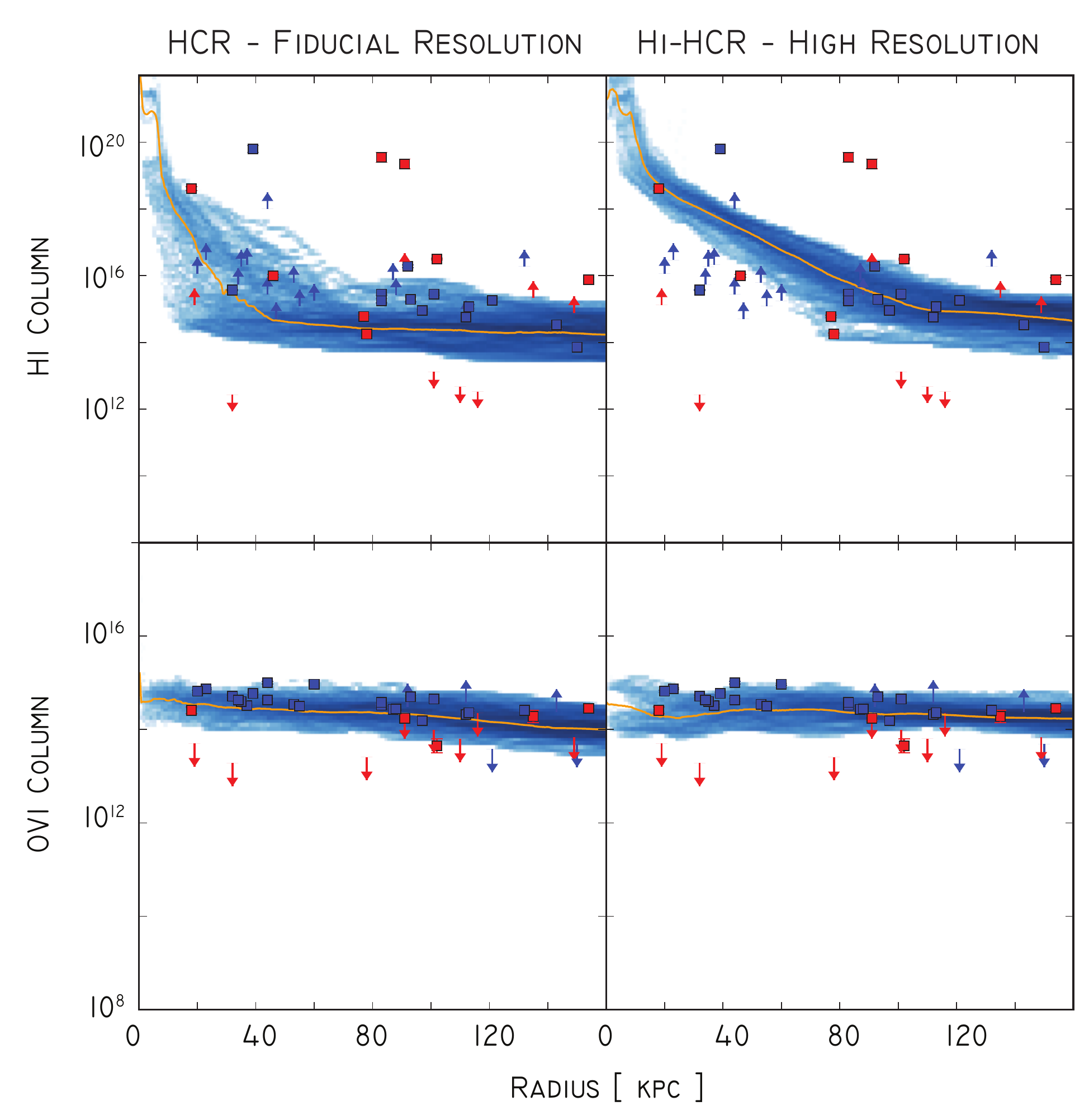}{A comparison of standard and high resolution runs of similar (though not identical) $M\approx 10^{12} M_\odot$ halos for the $\kappa_{\rm CR} = 3 \times 10^{28} \;  {\rm cm^2/s}$ CR model. Shown here is a comparison to COS-Halos, plotting column density versus impact parameter, as in Figure \ref{fig:cos-halos}. The left column plots the standard resolution \MCR\ run, with a $128^3$-cell base grid, while the right column comes from a zoom-simulation with a 256$^3$-cell base grid, and thus a factor of 8 higher mass resolution and two times higher spatial resolution. The high-resolution run features consistently higher HI column (top row) at low impact, but the same column towards 160 kpc. For OVI (bottom row) the two runs produce roughly identical distributions, both in good agreement with COS-halos.}{fig:hi-res}{.4}

\subsection{Resolution}
\label{sec:cgm-resolution}
Our simulations evolved on a base grid of resolution $128^3$ cells and an initial refinement region targeting our $\approx 10^{12} M_\odot$ halo of interest with two additional levels of refinement, each enhancing the resolution by a factor of two in each dimension. This implies a DM particle mass of $4.9 \times 10^6 M_\odot$. The base grid and initial, static refinements thus set the mass resolution of both DM and baryons. Seven additional levels of refinement were then adaptively introduced for the baryon fluid quantities, which set a maximum commoving spatial resolution of 305 $h^{-1}$ pc. In addition to better resolving the DM and baryonic substructure and dynamics, increased resolution, for fixed star formation parameters, tends to lead to higher star formation rates \citep{Tasker2006,Hummels2012,Salem2014a}.

To explore whether or not resolution strongly effects our CR-infused CGM, we performed additional high-resolution simulations for the $1$ and $3 \times 10^{28}$ cm$^2$/s diffusion models, with an initial base grid of $256^3$ cells and, as before, 2 levels of initial static refinement and 7 additional levels of AMR for the baryons. This lead to a DM particle mass of $6.1\times 10^5 M_\odot$ and a maximum spatial resolution of 153 $h^{-1}$ pc. Unlike the main body of work presented here, the initial conditions were generated with the multi-scale initial conditions program, or {\tt MUSIC} \citep{Hahn2011}. Thus these higher resolution results focus on a distinct halo with its own formation history and environmental properties (though also $\approx 10^{12} M_\odot$, relatively isolated with a rotationally supported disk). A detailed analysis of the higher resolution disk and CGM properties, resulting observables, and tests of convergence are all beyond the scope of this paper, and will appear in a follow-up (Salem \& Bryan 2015, in prep). For now, we note the results presented here are not substantially altered, though the corresponding \HCR\ run appears to harbor more cooler, HI gas in its inner CGM, more in line with the \MCR\ run presented here.

Figure \ref{fig:hi-res} presents a limited comparison of our \HCR\ run to the higher-resolution $10^{12} M_\odot$ halo we simulated with $\kappa_{\rm CR} = 3 \times 10^{28}$ cm$^2$/s, which we refer to as \HHCR. Here we again provide comparison to the COS-Halos data, as in Figure \ref{fig:cos-halos}, plotting ion column density versus impact parameter. For HI, \HHCR\ run features a consistently higher column than \HCR\ at low impact, due to an enhanced cavity of HI and generally cooler CGM material at low radii, as in the \MCR\ run. At larger radii, the two simulated distributions look roughly equivalent. For OVI, the distributions feature no significant disparities, again providing a solid match to the COS-Halos data for both late- and early-type systems across impact parameter.

\section{Summary}
\label{sec:cgm-conclusion}
In this paper we explored how cosmic rays can alter the structure and composition of the circumgalactic medium within a Milky Way-sized halo. We implemented a simple two-fluid model for the high energy CR proton population and the thermal ISM into \verb|Enzo|, an adaptive-mesh eulerian hydrodynamics code. We then followed the formation of a $10^{12} M_\odot$ halo from $z=99$ to the present day for five different models with and without CRs, implementing a simple prescription for radiative cooling, star formation and energetic feedback into both the thermal plasma and the diffusive CR fluid. The following list enumerates our key findings
\begin{enumerate}
\item Runs which included a diffusive CR component featured enriched CGMs, with gas metallicites $\sim 0.1 Z_\odot$ within a majority of the CGM volume out to beyond the virial radius. The enriched material is a direct consequence of robust winds flowing beyond the star forming disks within these runs. In contrast, a control run without CRs featured gas almost completely absent of metals in the CGM.
\item Beyond driving winds (and thus enriched material) into the galaxy's halo, the CR proton fluid also provided substantial pressure support within the CGM. For less-diffusive CR runs, this produced a spheroidal bubble of radius $r \lesssim \sqrt{t_{\rm dyn} \kappa_{\rm CR}}$ where the material was almost exclusively at $T < 10^{5}$ K, with a substantial presence of neutral hydrogen. This cooler material contributed roughly $40\%$ of the CGM's mass for a model with $\kappa_{\rm CR} = 10^{28}$ cm$^2/s$. This bubble's prominence diminished for a more diffusive run.
\item Beyond this inner region, the CR fluid provided the dominant source of pressure support in large swaths of the diffuse CGM out to the virial radius, while other, typically hotter, regions were gas pressure-dominated. This mixture led to a rich spread in temperatures that were on average measurably lower than in the non-CR run.
\item Diffusive cosmic ray runs also featured substantially stronger column densities of HI and important ions, such as CIII, SiIV and OVI, tracers of gas across many phases. For runs with $\kappa_{\rm CR} \in \{1,3\} \times 10^{28}$ cm$^{2}$/s, these columns provided a good match to the COS-halos survey of $L \sim L^*$ galaxies at $z = 0.2$.
\item The superior match to COS-halos achieved by the CR-inclusive runs was not solely due to enhanced CGM metallicity, via CR-driven winds from the disk. Rather, the cooler regions of diffuse CGM material afforded by CR pressure support was also a key ingredient in producing realistic ion columns, particularly for the lower energy species SiIV and CIII.
\item The CR-infused halo of our runs produced a luminous gamma-ray halo via hadronic (``catastrophic'') losses. This luminosity was strongly dependent on the diffusion coefficient, with a factor of three drop in the CR diffusion length-scale leading to over a factor of 10 rise in gamma-ray brightness. For the most diffusive runs, this brightness fell beneath the diffuse ``extra-galactic'' background found by Fermi LAT, and approached observed and expected values for M31 and the MW. For the less diffusive runs, this luminosity was far too high compared to observations, exceeding the CR energy output of the disk and thus rendering the models non-self consistent. For mildly diffusive runs, the effect of hadronic losses on the dynamics are an important process to model in future work.
\item These observations point towards a global CR diffusion coefficient near $3 \times 10^{28}$ cm$^2$/s, though a more accurate physical picture likely involves a spatially and time-varying diffusion coefficient, to reflect the non-linear interplay between gas, CRs and magentic fields, particularly far from the star-forming disk where the movement of ISM plasma may indeed comb out a more regularized field structure.
\item Cosmic ray protons likely alter the CGM's mass content, as broken down by temperature. In particular, less diffusive runs feature a high fraction of mass with $T \lesssim 10^5$ K, in line with recent results from the COS-halos survey. For more diffusive runs, however, gas mass at this low temperature diminishes in quantity.
\end{enumerate}

\section*{Acknowledgments}
We would like to thank Gurtina Besla, Alyson Brooks, Cameron Hummels, Mordecai-Mark Mac Low and Mary Putman for many useful discussion related to this work. We acknowledge financial support from NSF grants AST-0908390 and AST-1008134, and NASA grant NNX12AH41G,  as well as computational resources from NSF XSEDE, and Columbia University's Yeti cluster.

\bibliography{ms}{}


\section{Appendix: Uncertainty in the CR Spectrum}
\label{sec:cgm-spectrum}

To produce a $\gamma$-ray emissivity due to hadronic losses at each voxel in our simulations, we had to assume all CR protons lie above the energy threshold $q_{\rm thr} m_p c^2 = .78$ GeV, or a momentum threshold $q_{\rm thr} \approx 1$ in units of $m_p c$. For the well constrained CR spectrum of the solar neighborhood, this would overestimate the luminosity by over a factor of two, since over half the CR population's energy density comes from protons whose momentum lies below $m_p c$. Far beyond the star-forming disk of our simulated systems, where the CR population has not been replenished by SN, we would expect the contribution to $\epsilon_{\rm CR}$ from low-momentum protons to diminish, as lower-energy CR protons are preferentially susceptible to energy loss via Coulomb losses.

A more accurate approach would be to eliminate the contribution to $\epsilon_{\rm CR}$ from protons below the threshold $q_{\rm thr}$. However, our model does not explicitly track even an approximation of the CR momentum distribution. We can make a first pass at quantifying how strongly this will affect our results by considering the simple spectral model of \cite{Ensslin2007}
\begin{equation}
f(p) = C p^{-\alpha} \theta(p -q )
\label{eq:hadron}
\end{equation}
where $f(p)$ is the number density of CR protons of momentum $p$, $C$ is the normalization (amplitude) of the spectrum, $\alpha$ is a the spectral index, and $q$ is a low-momentum cutoff (distinct from $q_{\rm thr}$ for the hadron losses). The model's three parameters, $C$, $\alpha$ and $q$ are all functions of both space and time, and can thus capture the salient features of a variety of real CR populations. We can map this spectrum to the CR energy density via the integral 
\begin{eqnarray}
\epsilon_{\rm CR}(C,\alpha,q) &=& \int_0^\infty dp f(p) T_p(p)	\\
	&=&	\frac{Cm_pc^2}{\alpha-1} \times \left [ \frac{1}{2} \mathcal{B}_{\frac{1}{1+q^2}}\left( \frac{\alpha-2}{2}, \frac{3-\alpha}{2} \right) \right. \nonumber \\
	& & \left. +  q^{1-\alpha} \left( \sqrt{1+q^2} - 1  \right) \right]
\end{eqnarray}
where $T_p(p) = (\sqrt{1+p^2}-1)m_pc^2$ is the kinetic energy of protons of momentum $p$ and $\mathcal{B}_c(a,b)$ is the incomplete $\beta$-function.

\stdFig{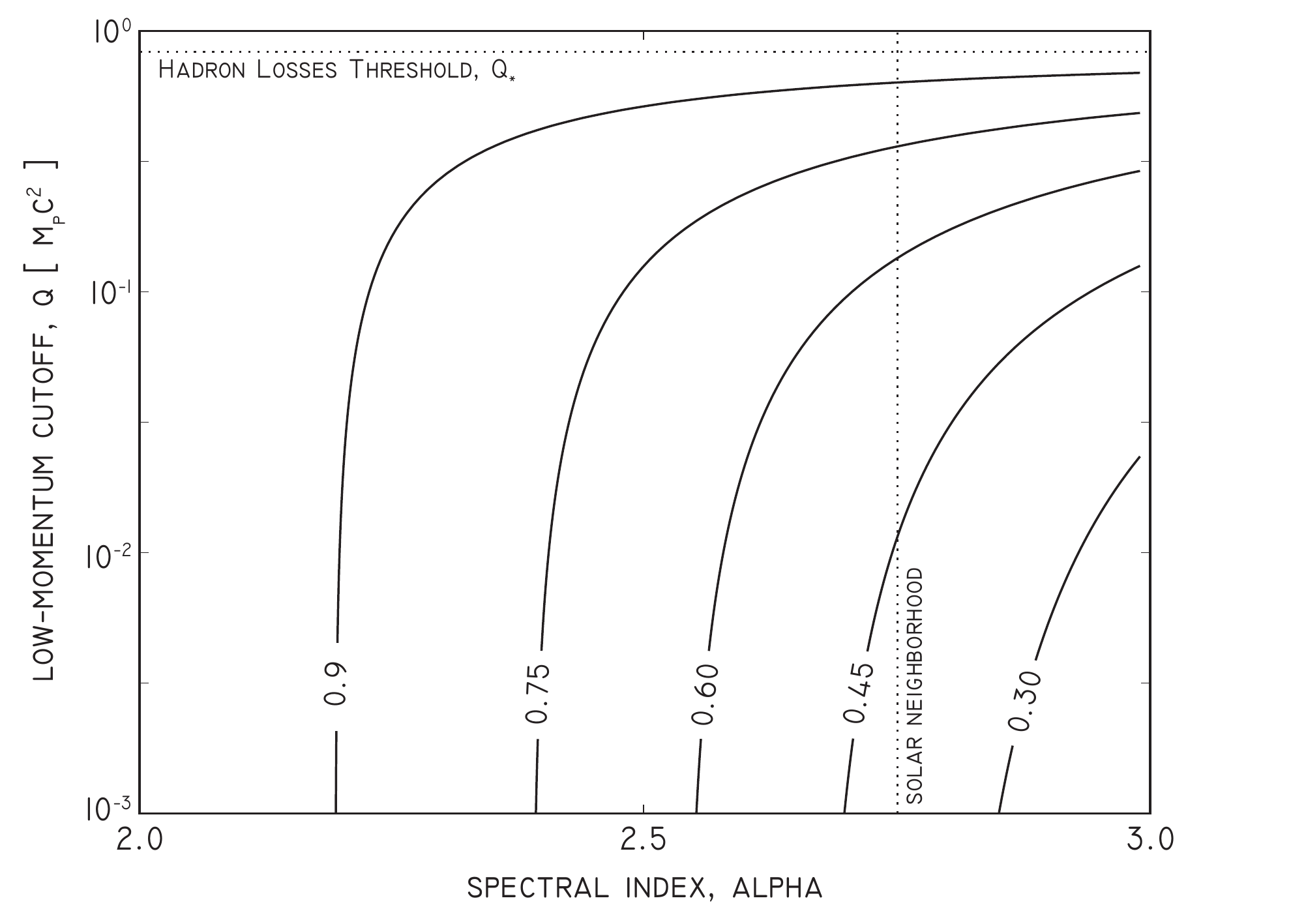}{The fraction of CR energy density that contributes to the hadron loss process, $f_{\rm had}$, as a function of the slope, $\alpha$, and cutoff, $q$, in the spectral model of Equation \ref{eq:hadron}. For the full range of $\alpha$ and reasonable values of $q$, this fraction is always $\in [0.3,0.9]$, and thus we find our results in Figure \ref{fig:hadron-luminosity} relatively insensitive to details of the CR spectrum. The vertical dotted line shows the slope value found in the solar neighborhood. The horizontal dotted line is the momentum cutoff for the hadron process, $q_\star$, above which $f_{\rm had}$ is unity.}{fig:param-space}{.45}

With this model, we can explore how inaccurate our assumption was in Section \ref{sec:cgm-hadron-luminosity} that all protons' momenta lie above $m_pc$. When the cutoff $q$ is \emph{above} $q_{\rm thr}$, this issue is moot: the CR gas is entirely composed of protons capable of producing $\gamma$-rays via hadronic losses. However, when $q$ drops below $q_{\rm thr}$, a fraction of the energy density should not be included in the product in Equation \ref{eq:hadron}. The relative importance of this cutoff issue will also be influenced by the slope coefficient, $\alpha$, as larger values will place more importance on lower momentum protons. We can express the fraction of the total energy density that participates as
\begin{equation}
f_{\rm had} = \frac{\epsilon(C,\alpha,q_{\rm thr})}{\epsilon(C,\alpha,q)} = f(\alpha,q)
\end{equation}
where the final right hand side of this equation emphasizes that the normalization $C$ drops out of this ratio, and thus the fraction is not explicitly dependent on the energy density at every point in space but rather only the slope and cutoff. Figure \ref{fig:param-space} shows $f_{\rm had}$ throughout this parameter space, demonstrating that the fraction is always $\in [0.3,0.9]$. This suggests details of the CR momentum distribution are unable to wash away the strong separation in the CGM's $\gamma$-ray luminosity across our simulations found in Figure \ref{fig:hadron-luminosity}. 

\end{document}